\documentclass[%
 reprint,
superscriptaddress,
 amsmath,amssymb,
 prx,
]{revtex4-2}

\usepackage{graphicx} % Required for inserting images
\usepackage{bm}% bold math
\usepackage{hyperref}% add hypertext capabilities
\usepackage{placeins}
\usepackage{physics}
\usepackage{babel}
\usepackage{xcolor}
\usepackage{dsfont}
\usepackage[caption=false]{subfig}
\usepackage{amsthm}
\usepackage{braket}
\usepackage{optidef}
\usepackage{algorithm2e}
\usepackage{tabularx}
\usepackage{multirow}
\usepackage{hhline}
\RestyleAlgo{ruled}

\makeatletter
\let\ORIbbl@fixname\bbl@fixname
\def\bbl@fixname#1{%
  \@ifundefined{languagealias@\expandafter\string#1}
    {\ORIbbl@fixname#1}
    {\edef\languagename{\@nameuse{languagealias@#1}}}%
}
\newcommand{\definelanguagealias}[2]{%
  \@namedef{languagealias@#1}{#2}%
}
\makeatother

\definelanguagealias{en}{english}
\definelanguagealias{EN}{english}
\definelanguagealias{en-GB}{english}

\newcommand{\btheta}{{\bm{\theta}}}
\newcommand{\bx}{{\bm{x}}}

\newcommand{\bphi}{\bm{\phi}}
\newcommand{\bPhi}{\bm{\Phi}}
\newcommand{\bTheta}{\bm{\Theta}}

\newcommand{\SU}[1]{\mathrm{SU}(#1)}
\newcommand{\su}[1]{\mathfrak{su}(#1)}

\usepackage{enumitem}
\newlist{abbrv}{description}{1}
\setlist[abbrv,1]{%
  labelwidth=2cm,
  labelsep=0.5cm,
  leftmargin=3cm,
  align=parleft,
  font=\normalfont,
  noitemsep,
  style=standard,
  format=\normalfont,
  wide=0pt,
  labelindent=0pt
}

\begin{document}

\title{Quantum Optimal Control with Geodesic Pulse Engineering}

\author{Dylan Lewis}
\affiliation{Department of Physics and Astronomy, University College London, London WC1E 6BT, United Kingdom}

\author{Roeland Wiersema}
\affiliation{Center for Computational Quantum Physics, Flatiron Institute, 162 Fifth Avenue, New York, NY 10010, USA}

\author{Sougato Bose}
\affiliation{Department of Physics and Astronomy, University College London, London WC1E 6BT, United Kingdom}
% \date{January 2025}

\begin{abstract}
Designing multi-qubit quantum logic gates with experimental constraints is an important problem in quantum computing. Here, we develop a new quantum optimal control algorithm for finding unitary transformations with constraints on the Hamiltonian. The algorithm, geodesic pulse engineering (GEOPE), uses differential programming and geodesics on the Riemannian manifold of $\SU{2^n}$ for $n$ qubits. We demonstrate significant improvements over the widely used gradient-based method, GRAPE, for designing multi-qubit quantum gates. Instead of a local gradient descent, the parameter updates of GEOPE are designed to follow the geodesic to the target unitary as closely as possible. We present numerical results that show that our algorithm converges significantly faster than GRAPE for a range of gates and can find solutions that are not accessible to GRAPE in a reasonable amount of time. The strength of the method is illustrtated with varied multi-qubit gates in 2D neutral Rydberg atom platforms. 
\end{abstract}

\maketitle

% \onecolumngrid

\section{Introduction}
% Quantum technologies promise to increase our capability to solve many computational problems~\cite{}. 
Quantum computing is underpinned by the precise manipulation of quantum systems. Quantum optimal control is a mature set of tools and methods, both analytical and numerical, to design control functions in quantum systems with experimental constraints. Optimal control theory predates quantum technologies and relates to optimising an objective function given a dynamical system with control parameters and constraints~\cite{goodwin2023, kallies2018}. Quantum optimal control involves applying many of these methods and theory to quantum systems~\cite{werschnik_quantum_2007, glaser_training_2015, boscain_introduction_2021}. The quantum systems are represented as Hamiltonians. The control parameters are typically time-dependent coefficients of 1- or 2-local Hamiltonian terms. A number of techniques are used to determine the control fields with distinct advantages and disadvantages~\cite{ansel_introduction_2024}. The most prominent methods are gradient-based, such as gradient ascent pulse engineering (GRAPE)~\cite{khaneja2005}. Other methods include: Krotov's method~\cite{krotov_global_1993, palao_optimal_2003, morzhin_krotov_2019}, which is also gradient based; chopped random basis (CRAB)~\cite{caneva_chopped_2011, muller_one_2022}, which is gradient free; and, more recently, a number of approaches that use machine learning~\cite{bukov_reinforcement_2018, day_glassy_2019, khalid_sample-efficient_2023, mao_machine-learning-inspired_2023} as well as stochastic optimal control methods \cite{lin2020stochastic, villanueva2024stochastic}. 
Quantum optimal control methods, and GRAPE in particular, are now regularly used to design qubit transformations on real quantum devices~\cite{koch2022}. GRAPE has been successfully used to solve many optimal control problems, including designing quantum gates for superconducting qubits~\cite{genois_quantum_2024}, NMR systems~\cite{khaneja_optimal_2005}, generating entanglement in spin chains~\cite{dolde_high-fidelity_2014}, mitigating noise~\cite{gorman_overcoming_2012}, and numerous other applications~\cite{koch2022}. 
% GRAPE is the optimal control method that we use to benchmark our method due to its broad use in many practical and theoretical applications.

In GRAPE, one has access to control parameters that determine the relative strength of an experimentally accessible Hamiltonian at each time step. Given a target unitary, one can calculate the trace distance (or fidelity) between the target and the controlled system. In the standard implementation GRAPE, the gradient of the fidelity is computed with respect to all control parameters concurrently and used to update the control parameters to maximally increase the fidelity~\cite{khaneja2005}, but second-order algorithms can be used to improve convergence~\cite{goodwin2023}. GRAPE is the mainstay, and indeed the representative of the current status of quantum optimal control.

Here, we introduce a completely different idea for a quantum optimal control algorithm.  We refer to this as geodesic pulse engineering (GEOPE). Rather than using gradients of the parameters, the GEOPE algorithm exploits the geodesic to the target unitary evolution on the special unitary manifold. Staying as close as possible to this geodesic while satisfying constraints makes this algorithm far more performant than existing techniques such as GRAPE. While our work is inspired by geometric methods for quantum circuit complexity and control~\cite{nielsen_quantum_2006, nielsen2006optimal,bhattacharyya2020renormalized,perrier2020quantum,carlini2007time,wang_quantum_2015, swaddle2017generating, swaddle2017thesis}, where varied concepts and tools such as sub-Riemannian manifolds, the quantum brachistochrone equation and machine learning have been deployed, it goes beyond and finds a practical (easy to implement) geometric method to obtain multi-qubit gates for more than $3$ qubits, which is still lacking.  
%which may, however, involve solving the quantum brachistochrone equation~\cite{carlini2007time}, which can be numerically challenging ~\cite{wang_quantum_2015, swaddle2017generating, swaddle2017thesis}.
%In previous work, using similar methods, an algorithm was introduced to find parameters for a single time-independent Hamiltonian to generate a target unitary~\cite{lewis_geodesic_2024}. 

% Our work falls in the category of geometric control methods~\cite{perrier2020quantum}, and is related to methods that make use of sub-Riemannian geometry~\cite{montgomery2002tour}, a connection that was originally proposed in the context of quantum control by Nielsen et al.~\cite{nielsen_quantum_2006, nielsen2006optimal}, where the complexity of implementing certain operations was related to finding the shortest geodesic on the unitary group. Such geodesics can in principle be found for constrained Hamiltonian dynamics by solving the quantum brachistochrone equation~\cite{carlini2007time}. However, this is a numerically challenging problem~\cite{wang_quantum_2015, swaddle2017generating, swaddle2017thesis}. 

In GEOPE, we only make use of the fact that the shortest geodesic generated by an unconstrained Hamiltonian can always be found straightforwardly through the principal branch of the logarithm. We then construct a linear problem that minimises the distance between this geodesic and the directions available in the constrained Hamiltonian. While we recently used a similar technique to find parameters for a single time-independent Hamiltonian to generate a target unitary~\cite{lewis_geodesic_2024}, that method is a far cry from optimal control and limited in scope in terms of versatility -- the time independence does not allow all gates under to be generated with realistic Hamiltonians. In fact, more importantly, Hamiltonian parameters can usually be varied in time, so it is important to optimally exploit the possibilities available, while remaining restricted to available hardware constraints for efficient multi-qubit gate design. %Although we do not have guarantees that our solution is a minimal geodesic on the sub-Riemannian manifold.
Our current algorithm can efficiently find solutions to the control problem for a very wide set of nontrivial multi-qubit gates. We compare GEOPE to first- and, more advanced, second-order GRAPE methods. GEOPE is found to significantly outperform GRAPE in the cases that we studied. We then use GEOPE to design up to a 6-qubit quantum Fourier transform, well beyond what is within reach for GRAPE. 

\section{\label{sec:algo}Geodesic pulse engineering} 
\subsection{Pulse Control}
We consider a quantum control setup where we have access to pulses generated by Hamiltonians of the form $H(\bphi) = \sum_k \phi_k G_k$, where $G_k$ is an $n$-qubit Pauli string and $\bphi$ is a set of control parameters. In practice, this Hamiltonian decomposes into a controllable part and a drift term
\begin{align}
    H = H_d + H_c,
\end{align}
where $H_d$ is the drift Hamiltonian with fixed parameters, and $H_c$ is the control Hamiltonian with adjustable degrees of freedom. In an experimental setup, $H$ only contains a small number of interactions, which correspond to the physical constraints of the system.
We denote this set of accessible interactions by $\mathcal{H}$, where we consider the case where $|\mathcal{H}|=O(n^2)$ for a typical 2-local Hamiltonian. 
Given a set of controls $\bPhi = (\bphi_1, \bphi_2, \ldots,\bphi_L)$, with $\phi_{l,k}$ the $k$th element of $\bphi_l$ and $k=1,\ldots,|\mathcal{H}|$, we can generate a sequence of $L$ pulses that generates a unitary
\begin{align}
    U_\textrm{G}(\bPhi) = U(\bphi_L) U(\bphi_{L-1}) ... U(\bphi_2) U(\bphi_1).
\end{align}
where $U(\bphi_l) = e^{iH(\bphi_l)}$. If the set of accessible directions in $\mathcal{H}$ generates the Lie algebra $\su{2^n}$ (controllability), $U_\textrm{G}(\bPhi)$ can approximate any target unitary given enough parameters and steps $L$~\cite{d2021introduction}.

We are interested in finding a set of controls $\bPhi$ such that $U_\textrm{G}(\bPhi)$ approximates $V$ up to some error $\varepsilon$  and a global phase. In particular, we consider the fidelity, 
\begin{align}
    F(\bPhi,V) = \frac{1}{N}\abs{\textrm{Tr}\{U_\textrm{G}^{\dagger}(\bPhi)V\}}\label{eq:fid}
\end{align}
where we defined $N=2^n$.
We consider $\bPhi$ a \textit{solution} if $F(\bPhi,V)>1-\varepsilon$. In GRAPE, the fidelity in Eq.~\eqref{eq:fid} is maximised with a gradient ascent method.

% GRAPE can be used to find the solution parameters for problems with a few qubits. In this work, we describe a new algorithm to solve the problem, GEOPE, which finds solutions much faster and for larger $n$. The algorithm is iterative, and the number of algorithmic steps is given by $m$, with a maximum number $M$ chosen to find a solution. For a sufficiently high number of piecewise steps in the evolution, $L$, the unitary can generate any desired unitary. As we will describe later in this section, when $L$ is high enough the algorithm will minimise an objective function for an underdetermined system of equations, meaning that there can be an infinite number of solutions. When $L$ is not high enough, minimising the objective function can be difficult. 

% In this section, we give an overview of the algorithm. A detailed description is given in App.~\ref{sec:algorithm}, which also explains the use of concepts from Riemannian geometry and differential programming.
\begin{figure}
    \centering
    \includegraphics[width=1.0\linewidth]{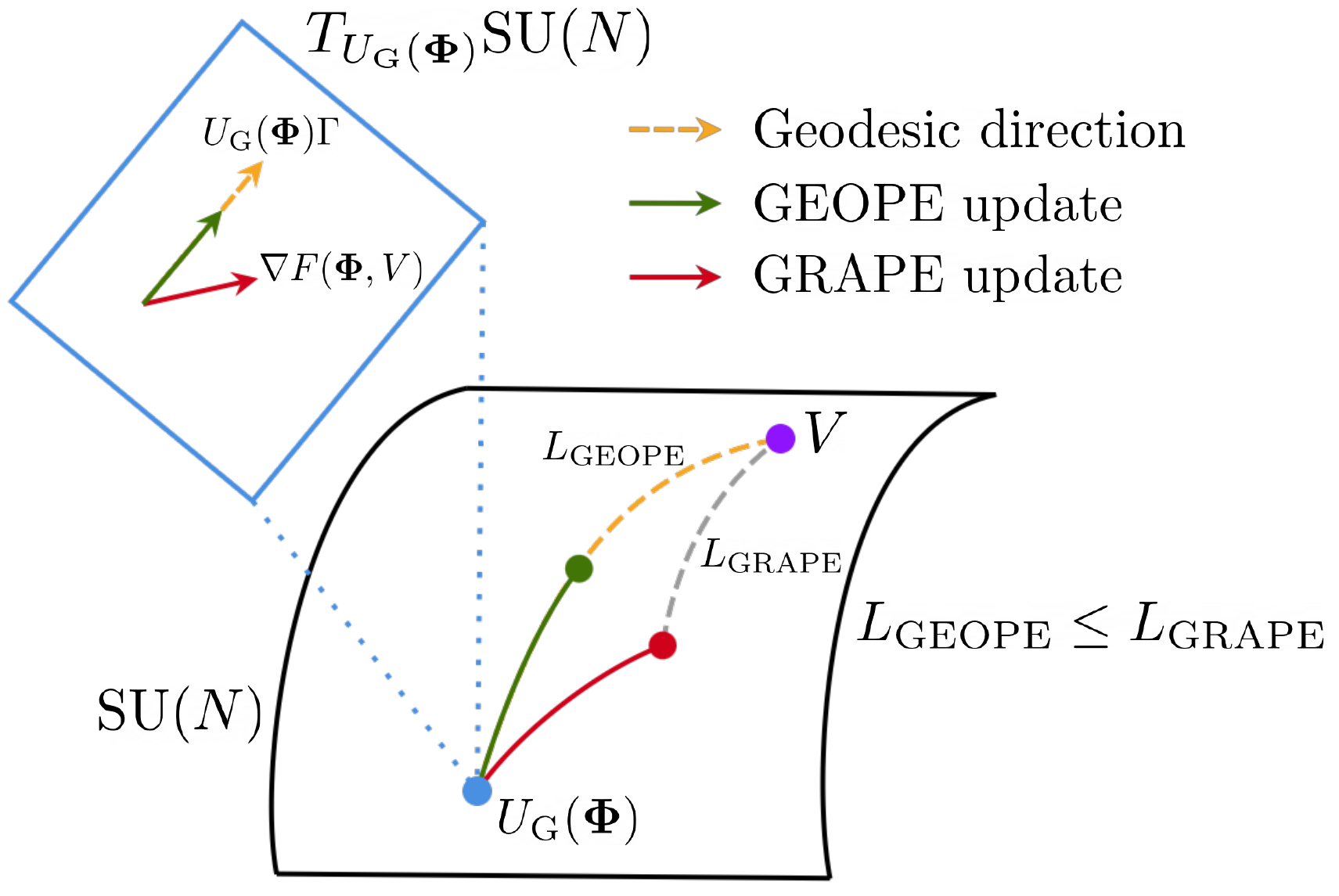}
    \caption{An illustration of a single algorithmic step of equal size for GEOPE and GRAPE. GEOPE steps in the direction of the geodesic $\Gamma$. GRAPE steps in the direction of maximum gradient of the fidelity, which does not necessarily align with the geodesic direction. $L_\textrm{GEOPE}$ and $L_\textrm{GRAPE}$ are the geodesic distances to the target $V$ after a step with GEOPE and GRAPE respectively. If the step sizes are equal, then the GEOPE update gets closer to the target on the manifold, $L_\textrm{GEOPE}\leq L_\textrm{GRAPE}$, with equality only if the direction of maximum fidelity is the geodesic direction. }
    \label{fig:geometry_grape_geope}
\end{figure} 

\subsection{Geodesic Optimisation}
Up to a global phase that we can neglect, unitary evolutions are elements of the special unitary group $U_\textrm{G}(\bPhi) \in \SU{N}$. We can identify a Hamiltonian as an element of the Lie algebra $H(\bphi_l) \in \su{N}$. Since Lie groups are smooth manifolds, we can equip $\SU{N}$ with a metric, $g(x,y) = \Tr\{x^\dagger y\}/N$ with $x$ and $y$ elements of the tangent space, to turn it into a Riemannian manifold. Due to the differentiable structure on $\SU{N}$, we can obtain the derivative at a point $U$ which is associated with a tangent space $T_U\SU{N}$. 

The core idea behind GEOPE is that we know the exact path from the current location on the manifold $U_\textrm{G}(\bPhi)$ to the solution $V$ at all times. This path is given by the one-parameter curve
\begin{align}
    X(t) &= U_\textrm{G}(\bPhi)\exp{it\Gamma}\nonumber\\
    \Gamma &= \log(U_\textrm{G}(\bPhi)^\dagger V) \in \su{N},
\end{align}
where we take the principal branch of the logarithm to ensure the shortest geodesic to the target. Note that $X(0)=U_\textrm{G}(\bPhi^{(m)})$ and $X(1) = V$. In practice, we only have access to a limited set of controls, which results in a restricted set of directions on $T_{U_G} \SU{N}$ that we can use to get closer to $V$.
This set of directions is given by the Jacobian
\begin{align}
    \mathbf{J}_{l,k}(\bPhi) = \frac{\partial U_G(\bPhi)}{\partial \phi_{l,k}} \in T_{U_G(\bPhi)} \SU{N}
\end{align}
with $l=1,\ldots,L$ and $k=1,\ldots,\abs{\mathcal{H}}$.
% The derivatives of the parameters at algorithmic step $m$ are represented by the Jacobian $\mathbf{J}_{l,k}(\bPhi^{(m)})$. The Jacobian exists in the tangent space, $T_{U_\textrm{G}(\bPhi^{(m)})}\SU{N}$. There is a geodesic from the current location on the manifold at algorithmic step $m$, $U_\textrm{G}(\bPhi^{(m)})$, to the target unitary, $V$. The geodesic, $X(t)$ is a curve on the manifold that is parameterised by a single parameter $t$ such that $X(0)=U_\textrm{G}(\bPhi^{(m)})$ and $X(1) = V$. Geodesics are also extrema of the path length functional. The geodesic is generated by a Lie algebra vector, $\Gamma$, which gives the direction from the current unitary to the target unitary. 
% The main idea in our algorithm is to update the parameters at each algorithmic step $m$ by $\bm{\delta\Phi}^{(m)}$ such that the resulting generator in the Lie algebra aligns the path of the geodesic $\Gamma^{(m)}$ as closely as possible. The \emph{update objective} is therefore given by 
\begin{figure*}[ht!]
    \centering
    \includegraphics[width=1\linewidth]{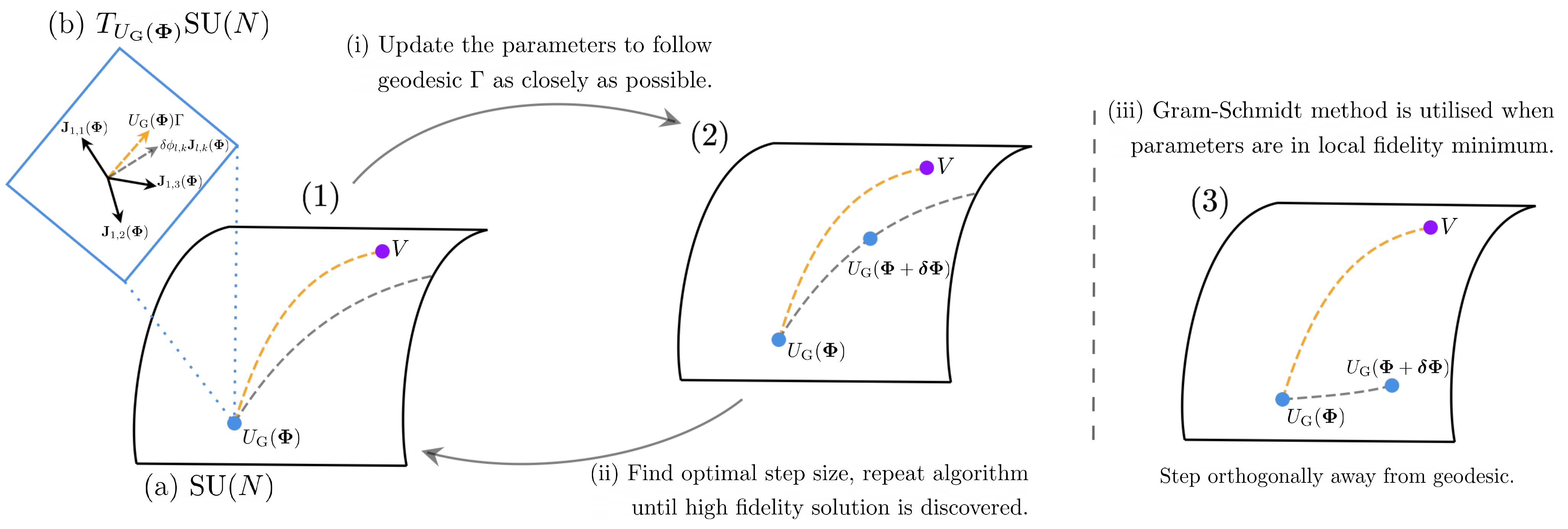}
    \caption{The labels (a) and (b) refer to depictions of the $\SU{N}$ manifold and its tangent space respectively. Only a few Jacobian matrices, $\mathbf{J}_{l,k}(\bPhi)$, have been included for clarity. At (1), we use the vectors of the Jacobians related to each parameter (black arrows) to find a linear combination (grey dashed arrow) that aligns with the generator of the geodesic (orange dashed arrow) as closely as possible. Having found the optimal direction, we proceed to (2), where a line search is performed over the direction to find the optimal step size. If the fidelity, $F(\bPhi+\bm{\delta\Phi},V)$, is increased, we return to (1) for the next algorithmic step. If the direction found in (1) cannot increase the fidelity for any step size up to the maximum step size, then we proceed to (3) and perform the Gram-Schmidt procedure before returning to (1). This is repeated until the maximum number of algorithm steps, $M$, or a solution is reached, which means a fidelity with error less than a predefined $\varepsilon$.}
    \label{fig:algorithm_schematic}
\end{figure*}
We now define the objective function that is minimised,
\begin{multline}
    \label{eq:update_condition}
    \mathcal{L}(\bm{\delta\Phi}) = \bigg\Vert \sum_{l=1}^{L} \sum_{k \in \mathcal{H}} \mathbf{J}_{l,k} (\bPhi) \delta\phi_{l,k} - i U_\textrm{G}(\bPhi) \Gamma \bigg\Vert^2,
\end{multline}
which measures how closely aligned a control update $\bm{\delta\Phi}$ is with the geodesic direction. Note that we multiplied $\Gamma$ with $U_G(\bPhi)$ to move the generator from the lie algebra $\su{N}$ to the tangent space $T_{U_G(\bPhi)}\SU{N}$.
If can obtain $\min\mathcal{L}=0$, the control update is perfectly aligned with the geodesic, hence we only need to perform a line search that maximises the fidelity to find the target $V$. If as a result of the limited number of controllable fields, $\min\mathcal{L}\neq 0$, the line search can fail to improve the fidelity, since we can only approximately follow the geodesic. 
In this case, we use a Gram-Schmidt procedure to move away from the local minimum -- this method, and all others for this algorithm, are detailed in App.~\ref{sec:algorithm}. A simplified schematic of the algorithm is shown in Fig.~\ref{fig:algorithm_schematic}.

In addition to the geometrical motivation of GEOPE, the cost function of Eq.\eqref{eq:update_condition} has two significant advantages over the one in Eq.~\eqref{eq:fid}. First, $\mathcal{L}(\bm{\delta\Phi})$ is convex and can be efficiently solved via a standard least-squares routine. 
% Second, $\mathbf{J}_{l,k}$ is $O(1)$, while the gradients of $F(\bPhi,V)$, which are required for GRAPE, vanish exponentially as $O(1/N)$. 
Furthermore, while it is believed that the control landscape should be trap free when the number of controls is much larger than $\dim(\su{2^n})$~\cite{rabitz2004quantum, rabitz2012comment}, in practice we are often working with a limited set of controls $L|\mathcal{H}|\ll \dim(\su{N})$ that can induce a complicated structure optimisation landscape with many local minima, which can greatly hinder the optimisation of Eq.\eqref{eq:fid}~\cite{tibbetts2012,ge2022optimization, beato2024towards}.

The algorithm presented here is a significant advancement of the single time-independent Hamiltonian case of Ref.~\cite{lewis_geodesic_2025} and establishes a method to solve the quantum optimal control problem for gate design.

% In general, GEOPE converges faster than GRAPE.
% Assuming that the step size of each algorithmic update on the special unitary manifold of both algorithms is set to be equal, the fastest possible algorithm would have update steps that follow a geodesic from the initial unitary $U_{\textrm{G}}(\bPhi)$ to the target unitary $V$, where the geodesic is the shortest path given by $X(t)$. 
The update steps of GEOPE follow the geodesic path as closely as possible at every algorithmic step, since this is the objective function that it directly minimises. 
% In App.~\ref{sec:geope_vs_grape}, we show that a small step along the geodesic path gives the maximum decrease in distance between $U_{\textrm{G}}(\bPhi)$ and $V$, as expected. 
In contrast, GRAPE uses the fidelity function of Eq.~\eqref{eq:fid} and finds the direction that locally maximises the fidelity. 
% In App.~\ref{sec:geope_vs_grape}, we find that the fidelity is locally maximised by taking a small step in the direction on the manifold $\mathbf{K}$ that maximises the expression
% \begin{align}
%         \label{sec:max_fid_expression}
%         \max_{\textbf{K}} \left[\frac{\textrm{Im}\left(\Tr{e^{i\Gamma}}^*\Tr{e^{i\Gamma} \textbf{K}}\right)}{\vert \Tr{e^{i\Gamma}}\vert} \right].
% \end{align}
% Although in some cases $\mathbf{K} = \Gamma$ may maximise the expression in Eq.~\eqref{sec:max_fid_expression}, this is \emph{not} the solution. In general, the geodesic direction is not the direction that maximises the local change in fidelity. 
% In App.~\ref{sec:geope_vs_grape}, we give a simple analytical example of $\mathbf{K} \ne \Gamma$. 
In App.~\ref{sec:geope_vs_grape}, we show that the update steps of GRAPE do not generally follow the geodesic and lead to a longer path to the solution, hence explaining why GRAPE can have slower convergence rates than GEOPE. This idea is illustrated in Fig.~\ref{fig:geometry_grape_geope}. Our numerical results in the following section demonstrate this clearly.
\begin{figure}[htb]
    \centering
    \includegraphics[width=0.9\linewidth]{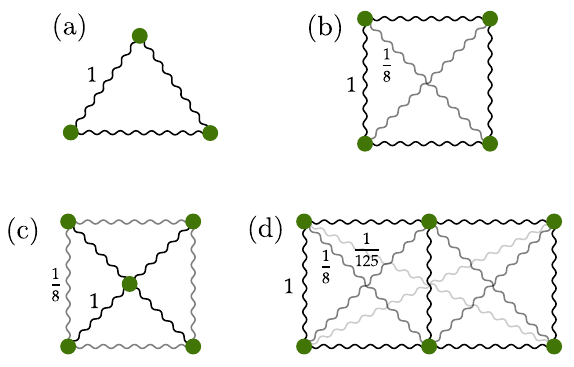}
    \caption{Graphs of the Rydberg interactions with the relative strengths, due to the $r^{-6}$ decay, of all the couplings shown for: (a) 3 Rydberg atoms, all interactions with equal strength; (b) 4 Rydberg atoms, the diagonal interactions have relative strength $\frac{1}{8}$ compared to the edge couplings; (c) 5 Rydberg atoms, the edge couplings have a relative strength of $\frac{1}{8}$ compared to the couplings to the central atom; (d) 6 Rydberg atoms, in addition to the 4 atom arrangement the long diagonal has a relative strength of $\frac{1}{125}$; 
    }
    \label{fig:rydberg_graphs}
\end{figure}

\section{\label{sec:results}Numerical results}
\begin{figure*}[tb!]
    \centering
    \subfloat[Toffoli, 12 piecewise steps]{\includegraphics[scale=0.61]{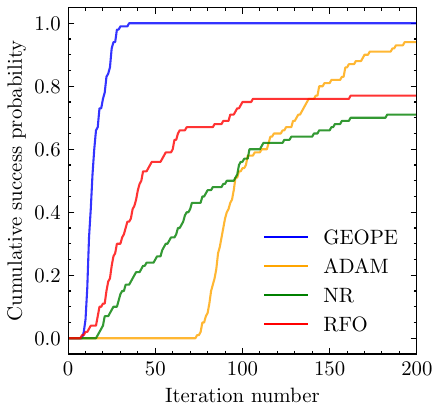}}
    \subfloat[CCZ, 12 piecewise steps]{\includegraphics[scale=0.61]{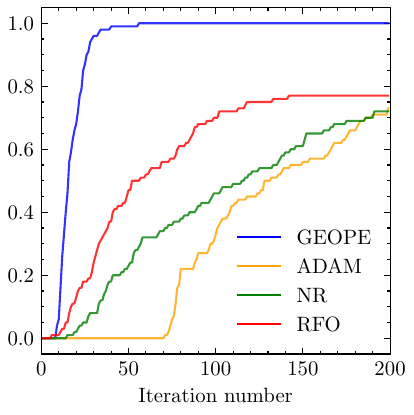}} 
    \subfloat[Toffoli, 20 piecewise steps]{\includegraphics[scale=0.61]{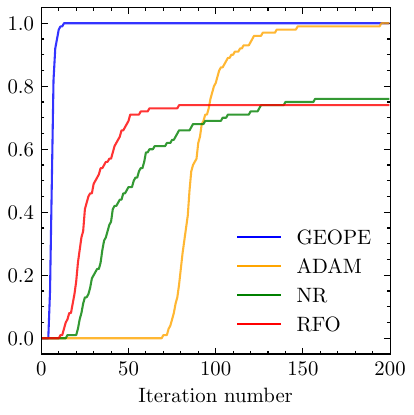}} 
    \subfloat[CCZ, 20 piecewise steps]{\includegraphics[scale=0.61]{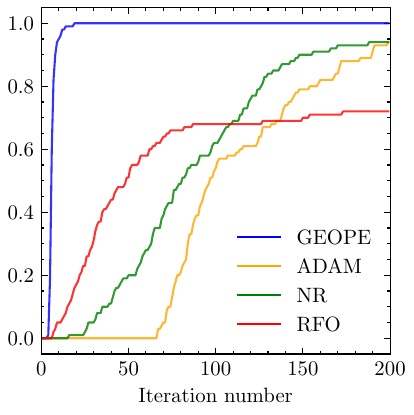}} 
    \caption{The cumulative success probability from 100 optimisation runs for finding: (a) CCX gate with 12 piecewise steps; (b) CCZ with 12 piecewise steps; (c) CCX with 20 piecewise steps; (d) CCZ with 20 piecewise steps. }
    \label{fig:cumulative_success}
\end{figure*}
Our method is general and works with any Hamiltonian restriction $\mathcal{H}$ and interaction topology. To demonstrate the power of GEOPE, we apply the method to a specific experimentally relevant case: using Rydberg atom arrays to generate multi-qubit quantum gates. 

The Hamiltonian available on such platforms  is given by~\cite{morgado_quantum_2021}
\begin{align}
    \label{eq:rydberg_atom_array_model}
    H(t) = \sum_{i<j}^n J_{ij} \sigma^z_i \sigma^z_j + \sum_{i=1}^n (\Omega_i(t) \sigma^x_i + \Delta_i(t) \sigma^z_i),
\end{align}
with coupling $J_{ij} \sim r^{-6}_{ij}$ where $r_{ij}$ is the distance between atoms $i$ and $j$~\cite{adams_rydberg_2019}. The interaction graphs for several 2-dimensional Rydberg lattices are shown in Fig.~\ref{fig:rydberg_graphs}.

To demonstrate the efficacy of our new quantum optimal control method, we compare our method to GRAPE and choose a quantum gate to find the pulses with the Rydberg atom array model of Eq.~\eqref{eq:rydberg_atom_array_model}. The simplest GRAPE method is a first-order optimisation and uses the Adam optimizer~\cite{kingma_adam_2017}. We also implement two second-order GRAPE methods, that use the Hessian to increase the convergence rate in terms of step number at the cost of additional computation. The second-order methods are the Newton-Raphson method (NR), and rational function optimisation (RFO)~\cite{goodwin_auxiliary_2015, goodwin_modified_2016}. All GRAPE methods are introduced and detailed in App.~\ref{sec:grape_methods} and our implementation of both GEOPE method and the reported GRAPE methods can be found in Ref.~\cite{our_data}. 

We now consider two target gates: a 3-qubit Toffoli and Controlled Z gate. We set $\varepsilon=10^{-9}$ and measure the cumulative probability of finding a solution for a given algorithmic iteration number. The plots in Fig.~\ref{fig:cumulative_success} demonstrate the efficacy of GEOPE, where the cumulative success probability of finding the two gates with the Rydberg control Hamiltonian of Eq.~\eqref{eq:rydberg_atom_array_model} is shown. The probabilities are experimentally found by performing the optimisation 100 times. For finding a Toffoli gate with 20 piecewise steps, GEOPE reaches a cumulative success probability of 1 within 13 algorithmic iterations. With the same number of steps, there is still no probability of success for GRAPE with Adam and Newton-Raphson methods, and around 0.05 cumulative success probability for GRAPE with RFO. The cumulative success probability reaches 1 for GRAPE with Adam after nearly 200 algorithmic iterations. In our experiments, we found that the significant improvement of GEOPE over GRAPE consistently applies for other gates and piecewise number of steps. Overall, the cumulative probability of success for GEOPE goes to 1 in at least 10 times fewer steps than GRAPE, when GRAPE finds a solution at all, for all the examples we tested. To ensure a fair comparison, we used Bayesian optimisation~\cite{bayesian_optimization_python} to find the optimal hyperparameters for each method, which is explained in detail in App.~\ref{sec:hyperparameters}.
\begin{figure}[htb]
    \centering
    \includegraphics[width=0.94\linewidth]{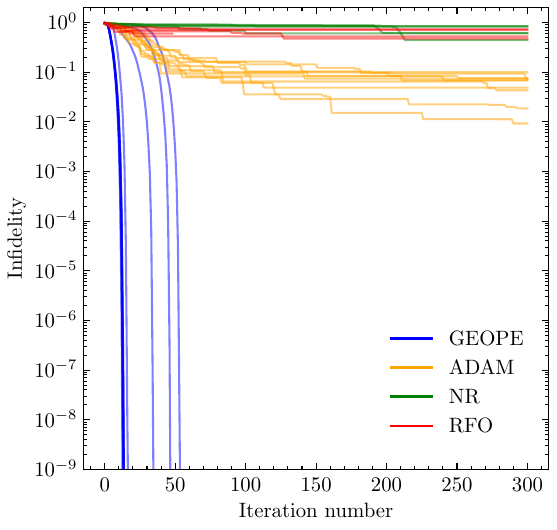}
    \caption{The infidelity of 10 optimisation trials for GEOPE and the GRAPE method is plotted against the algorithmic iteration number. 10 samples for GEOPE and Adam are plotted and 5 samples for the GRAPE methods Newton-Raphson and RFO. The target unitary gate is the 5-qubit QFT gate with 120 piecewise steps ($L=120$). The model is the Rydberg atom array of Eq.~\eqref{eq:rydberg_atom_array_model} with the interaction graph shown in Fig.~\ref{fig:rydberg_graphs}(c).}
    \label{fig:qft_5_method_comparison}
\end{figure}

For 5-qubit quantum gates, the disparity in success of the GEOPE and GRAPE methods becomes clear. In Fig.~\ref{fig:qft_5_method_comparison}, we design a 5-qubit QFT. In every trial, GEOPE quickly converges to a solution. GRAPE, on the other hand, does not find any solutions within 300 algorithmic steps. Although we do not claim that our implementation of these algorithms is optimal, we note that GEOPE finds these solutions in a couple of minutes, while the second-order GRAPE methods took hours to run (both using 16 CPU cores). Due to this excessive cost, which is mostly due to the calculation of the Hessian, we were unable to perform an extensive hyperparameter search and so
we reused the hyperparameters found for 3-qubit gates instead.

For the 6-qubit atom lattice given in Fig.~\ref{fig:rydberg_graphs}, we were able to find a 6-qubit QFT with GEOPE for $L=400$ (see Ref.~\cite{our_data} for the final parameter values), which was well beyond the capabilities of our GRAPE implementation.

\section{\label{sec:conclusion}Conclusion}
We have introduced a new quantum optimal control algorithm, GEOPE, that converges to solutions significantly faster than both first-order and second-order GRAPE and allows one to find large multi-qubit gates with a high probability of success. A result of GEOPE is unprecedented 5- and 6-qubit Quantum Fourier Transform gates in neutral Rydberg atom arrays via quantum optimal control. 

Our GEOPE algorithm is a first-order method, which means there are no second derivatives. On the other hand, the more advanced GRAPE methods use the Hessian to compute the step sizes~\cite{goodwin2023}. In principle, it is possible for our to incorporate the use of the Hessian into our method (see App.~\ref{sec:hessian}). However, we doubt that the additional computational cost and additional hyperparameter optimisation necessary is worth the potential improvement in convergence rate. The computational complexity for GEOPE is comparable to a first-order GRAPE method, with time complexity $O(KLN^4)$ rather than $O(KLN^3)$ for GRAPE, where $K=|\mathcal{H}|$ (see App.~\ref{sec:complexity}). 

Quantum optimal control is often concerned with preparing states rather than implementing unitary evolutions~\cite{bond_fast_2024}. GEOPE could most likely be used for this task as well, but this would require rephrasing the problem directly on the underlying Hilbert space or the related Homogenous space~\cite{wiersema2025horizontal}. 

GEOPE is flexible to any Hamiltonian restriction and can be applied with the Hamiltonian constraints that are relevant for ion traps, superconducting qubits, or semiconductor quantum dots. Future work could involve designing gates for these different platforms as well.

Currently, the piecewise step solutions for GEOPE and GRAPE in this work are square pulses that are not smoothed (see App.~\ref{sec:qft_solutions} for 3 and 4-qubit QFT solutions). In future work, the loss function $\mathcal{L}(\bm{\delta\Phi})$ can be modified to also minimise the distance between neighbouring piecewise pulses. This would allow solutions to be found with less severe changes in pulse magnitude, and continuous pulses could be found after smoothing. Continuous, slowly changing pulses are experimentally easier to implement and less prone to noise~\cite{nyisomeh_landauzener_2020, mcewen_removing_2021, ellert-beck_power-optimized_2024}. We can additionally add an energy constraint to minimise the total time $T$ of the evolution.

\section{Acknowledgements}	
DL acknowledges support from a UCL Quantum Proof of Concept grant, as well as a UCL Innovation and Enterprise Early Stage Commercialisation grant. DL, RW, and SB declare a relevant patent application: United Kingdom Patent Application No.~2400342.8. RW wants to thank Dries Sels for fruitful discussions during the preparation of this work. RW acknowledges support from the Flatiron Institute. The Flatiron Institute is a division of the Simons Foundation.

\clearpage

\onecolumngrid
% \appendix
\renewcommand{\thesection}{\Alph{section}}
\renewcommand{\theequation}{\Alph{section}\arabic{equation}}
\renewcommand{\thefigure}{\Alph{section}\arabic{figure}}
\renewcommand{\thetable}{\Alph{section}\arabic{table}}
\renewcommand{\thesubsection}{\Roman{subsection}} 
\counterwithin*{equation}{section}
\counterwithin*{table}{section}
\counterwithin*{figure}{section}
\setcounter{section}{0}
\pagebreak
\begin{center}
\textbf{\large Appendix}
\end{center}
%%%%%%%%%% Merge with supplemental materials %%%%%%%%%%
%%%%%%%%%% Prefix a "S" to all equations, figures, tables and reset the counter %%%%%%%%%%
\setcounter{equation}{0}
\setcounter{figure}{0}
\setcounter{table}{0}
\makeatletter

\section{\label{sec:notation}Notation}
We give a concise description of the symbols used. In the main text and the rest of the Appendix, the superscript $(m)$ indicating the algorithmic step number is often omitted for simplicity when it is not necessary.
\begin{table}[h!]
\renewcommand{\arraystretch}{1.25}
\begin{tabular}{p{1.6cm} p{14cm}}
$n$ & Number of qubits.\\
$N$ & Size of the Hilbert space, $N=2^n$.\\
$m$ & Denotes the algorithmic step iteration number.\\
$M$ & Denotes the maximum number of iterations.\\
$l$ & Denotes the piecewise step (layer) number. \\
$L$ & Denotes the maximum number of piecewise steps. \\
$k$ & Denotes the element of a Lie algebra vector. \\
$V$ & Target unitary transformation. \\
$U(\bm{x})$ & Unitary matrix generated by Lie algebra vector $\bm{x}$. \\
$U_\textrm{G}(\bm{X})$ & Product of $L$ unitary matrices, $U_\textrm{G}({\bm{X}}) = U(\bm{x}_L) U(\bm{x}_{L-1}) ...  U(\bm{x}_1)$ \\ 
$\btheta$ & A general unrestricted Lie algebra vector.\\
$\bphi$ & A general restricted Lie algebra vector.\\
$\btheta_l$ & Unrestricted Lie algebra vector of the $l$th piecewise step.\\
$\bphi_l$ & Restricted Lie algebra vector of the $l$th piecewise step.\\
$\theta_{l,k}$ & $k$th element of the unrestricted Lie algebra vector of the $l$th piecewise step.\\
$\phi_{l,k}$ & $k$th element of the restricted Lie algebra vector of the $l$th piecewise step.\\
$\bTheta$ & Matrix constructed from the $L$ unrestricted Lie algebra vectors $\btheta_l$, such that $\bTheta = (\btheta_1,\btheta_2, ..., \btheta_L)$ .\\
$\bPhi$ & Matrix constructed from the $L$ restricted Lie algebra vectors $\btheta_l$, such that $\bPhi= (\bphi_1,\bphi_2, ..., \bphi_L)$.\\
$\bPhi^{(m)}$ & Matrix of $L$ restricted Lie algebra vectors at algorithm step $m$. \\
$\bm{\delta}\bPhi^{(m)}$ & Change in parameters required to go from step $m$ to step $m+1$ in the matrix of parameter vectors, such that $\bPhi^{(m+1)} = \bPhi^{(m)} + \bm{\delta}\bPhi^{(m)}$ \\
$\Gamma^{(m)}$ & Generator of the geodesic path at algorithmic step $m$.\\
$\bm{\gamma}^{(m)}$ & Lie algebra vector for the generator of the geodesic path at $U_\mathrm{G}(\bPhi^{(m)})$.\\
$F(\bPhi, V)$ & Fidelity function for the unitary evolution of $\bPhi$ with the target unitary $V$.\\
$I(\bPhi, V)$ & Infidelity function defined as $1-F(\bPhi,V)$.\\
% $\Omega_{l,k}(\bPhi^{(m)})$ & Effective generator of piecewise step $l$ element $k$ for the restricted parameter matrix at algorithmic step $m$.\\
$\mathcal{H}$ & Set of Hamiltonian terms, the Lie algebra basis elements, that define the restriction.\\
$\mathcal{H}_\textrm{Rydberg}$ & Set of Hamiltonian terms, the Lie algebra basis elements, that define the restriction for the Rydberg Hamiltonian of Eq.~\eqref{eq:rydberg_atom_array_model}.\\
$R_\mathcal{H}$ & Matrix that projects an unrestricted Lie algebra vector to a restricted Lie algebra vector, $R_{\mathcal{H}} \btheta = \bphi$.\\
$\bm{R}_\mathcal{H}$ & Tensor that projects a matrix of unrestricted Lie algebra vectors to a matrix of restricted Lie algebra vectors, $\bm{R}_\mathcal{H} \bTheta = \bPhi$.\\
$\partial_{x_{l,k}}$ & Partial derivative with respect to entry ${l,k}$ of variable $x$. \\
$\mathbf{J}_{l,k}$ & Jacobian matrix of $U$, element $k$ of piecewise step $l$. \\ 
$\bm{j}_{l,k}$ & Lie algebra vectorisation of Jacobian matrix, element $k$ of piecewise step $l$. \\ 
$\bm{g}_{l,k}$ & Vectorisation of the gradient of the infidelity \\
$\mathbf{H}_{l,k,l',k'}$ & Matrix representation of the Hessian of the infidelity.\\
$\varepsilon$ & Maximum infidelity of a solution. \\
$\eta$ & Step size of the coefficients found by line search.\\
$\eta_\textrm{max}$ & Maximum step size, i.e the largest possible value of $\delta$. The key hyperparameter for GEOPE.\\ 
$\eta_\textrm{GS}$ & Step size of the coefficients for the Gram-Schmidt method. 
\end{tabular}
\end{table}
\clearpage
\section{\label{sec:geope}GEOPE}
In this section, we detail the GEOPE algorithm and then explain why it is faster at converging to solutions than GRAPE.
\subsection{\label{sec:algorithm}Algorithm}
Consider Hamiltonians acting on $n$ qubits as vectors in the Lie algebra $\su{2^n}$ with a basis of all the Pauli words, 
\begin{align}
    H(\btheta) = \sum_{j=1}^{4^n -1} \theta_j G_j = \btheta \cdot \boldsymbol{G},
\end{align}
where $G_j$ is the $j$th element of the set of Pauli words of length $n$, $\mathcal{P}_n$, ordered lexicographically,
\begin{align}
    \mathcal{P}_n = \{\sigma^{\alpha_1} \otimes \sigma^{\alpha_2} \otimes ... \otimes \sigma^{\alpha_n} \} \setminus \{\sigma^0 \otimes \sigma^0 \otimes ... \otimes \sigma^0 \}.
\end{align}
The Lie algebra vector $\btheta \in \mathbb{R}^{4^n -1}$ defines the Hamiltonian. In the same way, $\btheta$ with a time $t$ defines a unitary evolution
\begin{align}
    U(\btheta; t) = e^{i H(\btheta) t},
\end{align}
which is an element of the Lie group $\SU{2^n}$. However, we can always set $t=1$ and rescale $\btheta$, which is equivalent to rescaling the Hamiltonian strengths, giving $U(\btheta) = e^{i H(\btheta)}$. Lie groups are also smooth manifolds~\cite{helgason_differential_1979}. Equipping the smooth manifold with the metric $g(x,y) = \Tr{x^\dagger y}/N$ gives a Riemannian manifold. 

The parameters are restricted to certain Hamiltonian terms. Restricted vectors are given by $\bphi$. Restricted Hamiltonians are defined as Hamiltonians with non-zero coefficients only for certain Lie algebra basis terms. The set of Hamiltonian terms in the restricted is denoted by $\mathcal{H} \subset \mathcal{P}$. A general Lie algebra vector $\btheta$ can then be projected to a restricted vector by the diagonal matrix $R_\mathcal{H}$, with elements
\begin{align}
    (R_\mathcal{H})_{ij} = \delta_{ij} \bar{R}_\mathcal{H}(G_i),
\end{align}
where $\delta_{ij}$ is the Kronecker delta, $G_i$ is a basis element of the Lie algebra in set $\mathcal{P}_n$, and $\bar{R}_\mathcal{H}(G_i)$ is the binary function
\begin{align}
    \bar{R}_\mathcal{H}(G_l) = \begin{cases}
        1 & \textrm{if } G_l \textrm{ is in } \mathcal{H}, \\
        0 & \textrm{otherwise.}
    \end{cases}
\end{align}
Restrictions can significantly decrease the number of basis elements that need to be considered. For a $k$-local Hamiltonian, the number of terms grows polynomially as $O(n^k)$ with the number of qubits $n$ rather than exponentially as $O(4^n)$ for the full Lie algebra basis.

The quantum optimal control problem is that of finding the parameters for a tunable Hamiltonian that implements a desired overall unitary evolution, a quantum gate. The overall evolution $U_\textrm{G}$ consists of $L$ piecewise time-independent Hamiltonians to implement the gate,
\begin{align}
    U_\textrm{G}(\bPhi) = U(\bphi_L) U(\bphi_{L-1}) ... U(\bphi_2) U(\bphi_1),
\end{align}
where $\bPhi = (\bphi_1, ..., \bphi_L) \in \mathbb{R}^{(4^n-1)\times L}$ is the matrix of all piecewise parameters. The vector $\bphi_k \in \mathbb{R}^{4^n-1}$ has elements $\phi_{l,k}$. However, the restricted vectors $\phi_{l,k}$ are typically very sparse for restricted Hamiltonians -- a 2-local Hamiltonian would give a restricted vector $\bphi_{l,k}$ of rank $O(n^2)$ rather than $O(4^n)$ for the unrestricted vector. 

A geodesic algorithm was previously introduced that solved the problem of finding a single-step time-independent Hamiltonian that generates a desired unitary gate, i.e. when $L=1$~\cite{lewis_geodesic_2025}. However, for general unitary transformations, solutions cannot be found with a single time-independent Hamiltonian and $L>1$ would be required. Here, we present an new method to find the parameters for the experimentally important case of $L>1$. The algorithm can be applied to the standard optimal control setting. Our new geodesic algorithm for the quantum optimal control problem is compared to the most prominent existing approach, the gradient ascent pulse engineering method (GRAPE). Our method uses Riemannian geometry and differential programming to map the key update condition of the algorithm to a convex optimisation problem that can be solved efficiently.

The unitary gate $U_\mathrm{G}(\bPhi)$ depends smoothly on the parameters $\bPhi$ as a coordinate transformation, meaning there is a map from tangent vectors in $\mathbb{R}^{4^n -1}$ to tangent vectors in $\SU{2^n}$
\begin{equation}
    dU(\bm{x}) : T_{\bm{x}} \mathbb{R}^{4^n-1} \rightarrow T_{U(\bm{x})} \SU{2^n},
\end{equation}
with $\bm{x} \in \mathbb{R}^{4^n-1}$, having elements $x_{l}$. We note that the vectors $\bm{x}_k \in \mathbb{R}^{4^n-1}$ will have elements denoted by $x_{l,k}$.
This map is the Jacobian and can be found by differentiating the elements of the matrix $U(\bm{x}_k)$ with respect to all $x_{l,k}$. The JAX differential programming framework can provide this derivative numerically~\cite{jax2018github}. Computing the Jacobian at $\bPhi$ for $U_\textrm{G}(\bPhi) \in \SU{2^n}$ produces an element of the tangent space $T_{U_\mathrm{G}(\bPhi)} \SU{2^n}$. The elements of the Jacobian at $\bPhi$ are denoted by
\begin{align}
    (\mathbf{J}_{l,k} (\bPhi))_{i,j} = \frac{\partial}{\partial x_{l,k}} \left( U(\bm{x}_K) ... U(\bm{x}_1) \right)_{i,j} \bigg\vert_{(\bm{x}_1, ..., \bm{x}_K) = \bPhi}
\end{align}
where $(\mathbf{J}_{l,k} (\bPhi))_{i,j}$ is the $i,j$ element of the Jacobian matrix $\mathbf{J}_{l,k} (\bPhi)$ and $\left( U(\bphi_K) ... U(\bphi_1) \right)_{i,j}$ is the $i,j$ matrix element of $U_{\mathrm{G}}(\bPhi)$. 

The parameters at algorithm step $m$ can be updated for the subsequent step $m+1$. The change in parameters $\bm{\delta}\bPhi^{(m)}$ gives the new parameters $\bPhi^{(m+1)} = \bPhi^{(m)}+ \bm{\delta}\bPhi^{(m)}$, where $\bm{\delta\Phi}^{(m)} = (\bm{\delta\phi}_1,..., \bm{\delta\phi}_K) \in \mathbb{R}^{(4^n -1)\times K}$ and $\bm{\delta\phi}_k \in \mathbb{R}^{4^n-1}$ has elements $\delta\phi_{l,k}$. If the change in the parameters is small, the change to the unitary gate is well described by a Taylor expansion, 
\begin{align}
     U_\textrm{G}(\bPhi^{(m+1)}) = U_\textrm{G}(\bPhi^{(m)}) +  \sum_{k=1}^{K} \sum_{l \in \mathcal{H}} \mathbf{J}_{l,k} (\bPhi^{(m)}) \delta\phi^{(m)}_{l,k} + O(\delta\Phi^2),
\end{align}
where $\delta\Phi = |\bm{\delta}\bPhi^{(m)}|$. A small change in the parameter vectors is related to a small change on the special unitary group -- a change in $U_\textrm{G}(\bPhi)$. 

The problem is to find the parameters that generate a target unitary gate $V$. The question is how the unitary gate should be updated from $\bPhi^{(m)}$ to $\bPhi^{(m+1)}$ to get closer to $V$. We can construct the geodesic from the unitary at step $m$, $U_\textrm{G}(\bPhi^{(m)})$, to the target unitary, $V$. The geodesic path $X^{(m)}(t)$ in $\SU{2^n}$ is a one-parameter subgroup parameterised by $t$. At $t=0$, we require $U_\textrm{G}(\bPhi^{(m)}) X^{(m)}(0) = U_\textrm{G}(\bPhi^{(m)})$. At $t=1$, we require $U_\textrm{G}(\bPhi^{(m)}) X^{(m)}(1) = V$. The geodesic path is therefore $X^{(m)}(t) = e^{i\Gamma^{(m)} t} U_\textrm{G}(\bPhi^{(m)})$ with 
\begin{align}
    \Gamma^{(m)} = -i\log(U_\textrm{G}(\bPhi^{(m)})^\dagger V).
\end{align}
A small step of $\delta t$ along the geodesic direction updates the parameters as 
\begin{align}
    U_\textrm{G}(\bTheta^{(m+1)}) &= U_\textrm{G}(\bPhi^{(m)}) X^{(m)}(\delta t) \\
    &= U_\textrm{G}(\bPhi^{(m)}) + i U_\textrm{G}(\bPhi^{(m)}) \Gamma^{(m)} \delta t + O(\delta t^2). \nonumber
\end{align}
We want to update the parameters $\bPhi^{(m)}$ such that the tangent vector of the parameter update aligns with the geodesic as closely as possible. Updating the parameters in this way moves the unitary closer to the target unitary on the $\SU{N}$ manifold. The requirement for a small parameter update is that the change in parameters are in the direction of the geodesic to the solution as closely as possible. This requirement gives the \emph{update objective},
\begin{align}
    \mathcal{L}(\bm{\delta\Phi}^{(m)}) =  \bigg\Vert \sum_{l=1}^{L} \sum_{k \in \mathcal{H}} \mathbf{J}_{l,k} (\bPhi^{(m)}) \delta\phi^{(m)}_{l,k} - i U_\textrm{G}(\bPhi^{(m)}) \Gamma^{(m)} \bigg\Vert,
\end{align}
which is to be minimised with respect to $\bm{\delta\Phi}^{(m)}$.
The $\delta t$ has been neglected since it is a scalar that can absorbed into the small step $\delta\phi^{(m)}_{l,k}$. 
The geodesic algorithm for optimal control aims to satisfy this requirement by choosing the optimal update parameters $\delta\phi^{(m)}_{l,k}$ at each step $m$ to move closer to the desired unitary gate $V$. The update has become a linear least-squares problem, which is a convex optimisation problem. 
To solve the convex optimisation problem, we first map the problem to a Euclidean vector space by defining the vectors
\begin{align}
    \bm{j}_{l,k} = \sum_{j=1}^{N^2 -1} \Tr{G_j \mathbf{J}_{l,k}(\bPhi)} \hat{e}_j,
\end{align}
and 
\begin{align}
    \bm{\gamma} = \sum_{j=1}^{N^2 -1} \Tr{G_j U_\textrm{G}(\bPhi)\Gamma(\bPhi)} \hat{e}_j,
\end{align}
where $G_j$ are the Lie algebra basis elements of the ordered set $\mathcal{P}$, and $\hat{e}_j$ are unit vectors of the Euclidean vector space. The least squares problem now becomes the minimisation of the objective function
\begin{align}
    L(\bm{\delta\Phi}^{(m)}) = \bigg\Vert \sum_{l=1}^{L} \sum_{j \forall G_j \in \mathcal{H}} \left[ \delta\phi_{l,j}^{(m)} \bm{j}_{l,j} \right] - \bm{\gamma}\bigg\Vert^2.
\end{align}
Many terms in $\bm{\delta\phi}^{(m)}_l$ are constrained to be $0$ due to the restriction $R_\mathcal{H}$. The number of parameters to optimise over is therefore $L|\mathcal{H}|$ where $L$, as previously defined, is the total number of layers and $|\mathcal{H}|$ is the cardinality of the set restriction set $\mathcal{H}$.

The minimisation of the objective function gives the direction of the update step for $\bm{\delta\Phi}^{(m)}$. The magnitude of the step, $\eta$, is then determined by a golden section line search with a maximum step size given by a hyperparameter $\eta_\textrm{max}$, the line search is a maximisation of the \emph{fidelity} function 
\begin{align}
    F(\bPhi^{(m)},V) = \frac{1}{N}\Tr{U^\dagger_G(\bPhi^{(m)}) V},
\end{align}
where $V$ is the target unitary evolution. The \emph{infidelity} is similarly defined as $I(\bPhi, V) = 1-F(\bPhi,V)$. It is possible that the line search does not find a step magnitude $\eta$ that improves the fidelity, $F(\bPhi^{(m)}, V) > F(\bPhi^{(m)}+\eta\overline{\bm{\delta\Phi}}^{(m)}, V)$, where $\overline{\bm{\delta\Phi}}$ is the normalised $\bm{\delta\Phi}$, such that $\eta$ is the magnitude. In this case, we use a Gram-Schmidt method to step in a direction orthogonal to the geodesic to the solution. This minimises the chance that the algorithm steps back into the same minimum at the next step. The method uses $L$ random restriction vectors $\bm{r}_\mathcal{H},l$. The Gram-Schmidt update direction is therefore
\begin{align}
    \bm{\delta\phi}_l = \bm{r}_{\mathcal{H}, l} - \frac{\bm{r}_{\mathcal{H}, l} \cdot \bm{\gamma}}{\Vert \bm{\gamma} \Vert},
\end{align}
for each $l$, giving $\bm{\delta\Phi} = (\bm{\delta\phi}_1, \bm{\delta\phi}_2, ..., \bm{\delta\phi}_L)$. The Gram-Schmidt step size is a hyperparameter that determines the magnitude of the algorithmic step. Our experiments found that the Gram-Schmidt step size is more effective when it is larger than the maximum step size. We choose $\eta_\textrm{GS} = \Vert \bm{\delta\Phi} \Vert = 1.2 \eta $ for the Gram-Schmidt step size, where $\eta_\textrm{max}$ is the maximum step size. 

Algorithm~\ref{alg:unitary_gate_design} details the geodesic pulse engineering (GEOPE) method. In order to calculate $dU_{l,k}$ effectively, we make use of the observation in Ref.~\cite{goodwin_auxiliary_2015} that
\begin{align}
\left(
    \begin{array}{cc}
        U(\bx_l) & \partial_{x_{l,k}} U(\bx_l)\\
        0 & U(\bx_l)
    \end{array}\right)
    =
    \exp \left[i\times\left(
    \begin{array}{cc}
        \bx_l\cdot \boldsymbol{G} & G_k\\
        0 & \bx_l\cdot \boldsymbol{G}
    \end{array}
    \right)
    \right]\label{eq:aux_jac}
\end{align}

which allows us to calculate the partial derivative $\partial_{x_{l,k}}U(\bx_l)$ by calculating the matrix exponential of a $2\times 2$ block matrix containing $2^n\times 2^n$ matrices. The \texttt{Minimise} subroutine solves the least-squares problem using a standard singular value decomposition. The \texttt{Maximise} subroutine corresponds to a golden section line search to find the optimal step size $\eta$, given a maximum step size $\eta_\textrm{max}$.
\begin{algorithm}
\caption{Geodesic pulse engineering (GEOPE).}\label{alg:unitary_gate_design}
\KwIn{$V$, $\bm{\Theta}, R_{\mathcal{H}},  \varepsilon, \eta_\textrm{max},\eta_{\textrm{GS}},  M$}
\KwOut{$\bm{\Phi}$}
\text{Obtain the Jacobian function:}\\
\For{$l \in (1,\ldots, N^2-1$)}{
\For{$k \in (1,\ldots, K)$}{
$dU_{l,k}(\bx) = \partial_{x_{l,k}} \mathfrak{Re}[U(\bx_K)...U(\bx_1)] + i \partial_{x_{l,k}}\mathfrak{Im}[U(\bx_K)...U(\bx_1)]$}}

\text{Define evolution:} $U_G(\bTheta) = U(\btheta_K)...U(\btheta_1)$

\text{Define a fidelity function:} $F(\bTheta, V) = \Tr{U_G(\bTheta)^\dagger V}/N$

\text{Perform the Hamiltonian restriction:} $\bPhi \gets \bm{R}_{\mathcal{H}}\bTheta$

\For{$m \in \{ 1, 2, ..., M\}$}{
    \text{Compute the Jacobian at $\bPhi$:}\\
    {$\mathbf{J}_{l,k}(\bPhi) \gets dU_{l,k}(\bx)|_{\bPhi}$
    
    $\bm{j}_{l,k} \gets \sum_{j=1}^{N^2 -1} \Tr{G_j \mathbf{J}_{l,k}(\bPhi)} \hat{e}_j$}
    
    \text{Find the geodesic at $U_\textrm{G}(\bPhi)$:}\\
    $\Gamma \gets -i \log(U_G(\bPhi)^\dagger V)$\\
    $\bm{\gamma} \gets \sum_{j=1}^{N^2 -1} \Tr{G_j U_\textrm{G}(\bPhi)\Gamma(\bPhi)} \hat{e}_j$ 

    \FuncSty{Minimise}$\left(\Big\Vert \sum_{l=1}^{L} \sum_{j \forall G_j \in \mathcal{H}} \left[ \delta\phi_{l,j} \bm{j}_{l,j} \right] - \bm{\gamma}\Big\Vert^2\right) \textrm{ for normalised } \bm{\delta\bPhi}$

    \FuncSty{Maximise}$\left(F(\bPhi + \eta \bm{\delta\Phi}, V)\right) \textrm{ for } \eta$ with $\eta \leq \eta_\textrm{max}$
    
    \uIf{$F(\bPhi + \eta\bm{\delta\Phi}, V) < F(\bPhi, V)$}{choose $L$ random $\bm{r_l}$

    $\bm{r}_{\mathcal{H},l} \gets R_\mathcal{H}\bm{r}_l$

    \ForEach{l}{ 
    $\bm{\delta\phi}_l \gets \bm{r}_{\mathcal{H},l} - \frac{\bm{r}_{\mathcal{H},l}\cdot\bm{\gamma}}{\Vert\bm{\gamma} \Vert^2} \bm{\gamma}$
    }
    $\bPhi \gets \bPhi + \eta_{\textrm{GS}} \bm{\delta\Phi}$
    }
    \Else{
    $\bPhi \gets \bPhi + \eta \bm{\delta\Phi}$
    }
    \uIf{$F(\bPhi, V) > 1-\varepsilon$}{
    break
    }
}
\end{algorithm}

\subsection{\label{sec:geope_vs_grape}Why GEOPE converges faster than GRAPE}
The update steps of GEOPE follow the geodesic path as closely as possible at every algorithmic step, since this is the objective function that it directly minimises. 
In this section, we show that a small step along the geodesic path gives the maximum decrease in distance between $U_{\textrm{G}}(\bPhi)$ and $V$, as expected. 
In contrast, GRAPE uses the fidelity function of Eq.~\eqref{eq:fid} and finds the direction that locally maximises the fidelity. 
Although in some cases the geodesic direction may maximise the increase in expression in fidelity, this is \emph{not} the solution. We show that, in general, the geodesic direction is not the direction that maximises the local change in fidelity. 
We give a simple analytical example of $\mathbf{K} \ne \Gamma$. The result of this is that GRAPE will follow a longer path to the solution and lead to slower convergence rates when compared to GEOPE. 

Increasing the fidelity between $U_\textrm{G}(\bPhi)$ and $V$ is correlated to decreasing the geodesic distance between them but it is not strictly aligned. The geodesic distance in $\SU{N}$ is the path distance along the geodesic. The geodesic curve, as stated before, is $X(t) = U_\textrm{G}(\bPhi) \exp{i t \Gamma } $, with $\Gamma = -i \log( U_\textrm{G}(\bPhi)^\dagger V )$. The length of the geodesic path up to $\tau \in [0,1]$, with the metric as previously defined, is 
\begin{align}
    L[X(t), \tau] &= \int_0^\tau \sqrt{g(\dot{X}(t),\dot{X}(t))} dt \\
    &= \int_0^\tau \sqrt{\Tr{(\dot{X}(t)^\dagger \dot{X}(t)}} dt,
\end{align}
where $\dot{X}(t) = \frac{dX(t)}{dt}$. We have 
\begin{align}
    \dot{X}(t) &=  U_\textrm{G}(\bPhi) \frac{d}{dt} \left( \exp{i t \log( V U_\textrm{G}(\bPhi)^\dagger)} \right)  \\
    &= i  U_\textrm{G}(\bPhi) \Gamma \exp{i t \Gamma} \\
    &= i X(t) \Gamma.
\end{align}
This gives the path length
\begin{align}
    L[X(t), \tau] &= \int_0^\tau \sqrt{\Tr{\Gamma^\dagger X(t)^\dagger  X(t) \Gamma}} dt \\
    &= \int_0^\tau \sqrt{\Tr{ \Gamma^\dagger \Gamma }} dt \\
    &= \sqrt{\Tr{ \Gamma^2 }}  \tau.
\end{align}
If the geodesic has Lie algebra vector $\bm{\gamma}$, the geodesic path length is 
\begin{align}
    L[X(t), 1] = \Vert \bm{\gamma} \Vert. 
\end{align}

The next question is how a small deviation of $U_\textrm{G}(\bPhi)$ along the tangent space element $\textbf{J}$, with Lie algebra vector $\bm{j}$, changes the geodesic path length,
\begin{align}
    U_\textrm{G}(\bPhi^\prime) = U_\textrm{G}(\bPhi) \exp{i \epsilon \textbf{K}}, 
\end{align}
where we consider a small $\epsilon$ and $\textbf{K} = U_\textrm{G}(\bPhi)^\dagger \textbf{J}$. We now have $X^\prime(t) = U_\textrm{G}(\bPhi) \exp{i \epsilon \textbf{K}}  \exp{i t \Gamma^\prime}$, with $\Gamma^\prime = -i\log(\exp{-i\epsilon \textbf{K}} U_\textrm{G}(\bPhi)^\dagger  V )$. For small $\epsilon$, we can consider the Taylor series expansion 
\begin{align}
    \Gamma^\prime &= -i\log( U_\textrm{G}(\bPhi)^\dagger  V ) -i \epsilon \frac{d}{d\epsilon} \log(\exp{-i\epsilon \textbf{K}} U_\textrm{G}(\bPhi)^\dagger  V ) \Big\vert_{\epsilon=0} + O(\epsilon^2),
\end{align}
defining $W = U_\textrm{G}(\bPhi)^\dagger  V$, and using the Fr\'{e}chet derivative of the logarithm gives~\cite[Theorem 6]{haber_notes_nodate},
\begin{align}
    \Gamma^\prime = \Gamma - \epsilon \int_0^{1} [sW + (1-s)I]^{-1}  \textbf{K} W [sW + (1-s)I]^{-1}  ds + O(\epsilon^2).
\end{align}
We define 
\begin{align}
    D(\bPhi,\textbf{K}) = \int_0^{1} [sW + (1-s)I]^{-1}  \textbf{K} W [sW + (1-s)I]^{-1}  ds,
\end{align}
where $D(\bPhi,\textbf{J})$ is also Hermitian.
This gives a new geodesic path length 
\begin{align}
    L[X^\prime(t), 1] &= \sqrt{\Tr{\Gamma^2 - \epsilon\{D(\bPhi, \textbf{K}), \Gamma\} + O(\epsilon^2)} } \\
    &= \sqrt{\Tr{\Gamma^2}} - \frac{ \epsilon \Tr{D(\bPhi, \textbf{K}) \Gamma}}{\sqrt{\Tr{\Gamma^2}}} + O(\epsilon^2),
\end{align}
where $\{.,.\}$ is the anticommutator.
which can be written in terms of the previous geodesic path length
\begin{align}
     L[X^\prime(t), 1] = L[X(t), 1]- \frac{ \epsilon \Tr{D(\bPhi, \textbf{K}) \Gamma}}{L[X(t), 1]} + O(\epsilon^2),
\end{align}
for small $\epsilon$. If $\textbf{K}$ is along the same direction as $\Gamma$, so $\textbf{K} = \Gamma$, we have $[\mathbf{K}, W] = [\Gamma, W]=0$. Evaluating the first-order term gives
\begin{align}
    D(\bPhi,\textbf{K}) &= \textbf{K} W \int_0^{1} [sW + (1-s)I]^{-2}     ds \\
    &= \textbf{K} W W^{-1} \\
    &= \textbf{K} = \Gamma
\end{align}
Thus, we find a change in the geodesic distance of
\begin{align}
    L[X^\prime(t), 1] &= L[X(t), 1]-  \frac{ \epsilon \Tr{ \Gamma^2}}{L[X(t), 1]} + O(\epsilon^2), \\
    &= \Vert \bm{\gamma}\Vert (1-\epsilon).
\end{align}
As expected, moving in the geodesic direction decreases the geodesic distance by the distance moved $\epsilon$. This is the maximum reduction in distance between $U_\textrm{G}(\bPhi)$ and $V$ possible with a curve of path length $\epsilon \Vert \bm{\gamma} \Vert$. GEOPE will follow this direction as closely as possible to minimise the geodesic distance.

Now a small deviation of the unitary $U_\textrm{G}(\bPhi)$ along the tangent vector $\textbf{J}$ will also alter the fidelity. As before $W = U_\textrm{G}(\bPhi)^\dagger V$, and we have
\begin{align}
    F(\bPhi^\prime, V) &= \frac{1}{N} \big\vert\Tr{ W \exp{-i\epsilon\textbf{K}}}\big\vert \\
    &= \frac{1}{N}\big\vert\Tr{ W } - i\epsilon\Tr{W \textbf{K} } + O(\epsilon^2)\big\vert\\
    &= F(\bPhi, V) + \frac{\epsilon}{N} f_1 + O(\epsilon^2) 
\end{align}
The first order term is 
\begin{align}
    f_1 = \frac{1}{\vert \Tr{W}\vert} \textrm{Im}\left( \Tr{W}^* {\Tr{W\textbf{K}}} \right).
\end{align}
The maximum increase in fidelity for small $\epsilon$ is when $f_1$ is maximised. 
Using the geodesic direction, $V = U_\textrm{G}(\bPhi) \exp{i\Gamma} $, gives $W = \exp{i\Gamma}$. The maximum fidelity increase occurs at the $\textbf{K}$ that gives 
\begin{align}
\label{eq:max_f1}
    \max_{\textbf{K}} \left[\frac{\textrm{Im}\left(\Tr{e^{i\Gamma}}^*\Tr{e^{i\Gamma} \textbf{K}}\right)}{\vert \Tr{e^{i\Gamma}}\vert} \right].
\end{align}
The key point is that this expression is \emph{not} generally solved by the $\textbf{K} = \Gamma$. In fact, the geodesic direction is typically not the direction that maximises the change in fidelity. A simple example can demonstrate this observation. Take a 2-qubit system where the geodesic to the target unitary is the tangent space vector
\begin{align}
    \Gamma = \sigma^x_1 \sigma^x_2 + \sigma^y_1 \sigma^y_2 + \sigma^z_2,
\end{align}
and the initial vector $U(\bPhi)$ is the identity, so $\bPhi = \bm{0}$. The geodesic vector can be exponentiated to give a closed form, 
\begin{multline}
    e^{i\Gamma} = 
    \frac{I}{2} \left( \cos(1) +  \cos(\sqrt{5}) \right) 
    + i \frac{\sigma^z_1}{2} \left( \sin(1) - \frac{\sin(\sqrt{5})}{\sqrt{5}}\right) 
    + i \frac{\sigma^z_2}{2} \left( \sin(1) +  \frac{\sin(\sqrt{5})}{\sqrt{5}}\right) \\
    + \frac{\sigma^z_1 \sigma^z_2}{2} \left( \cos(1)- \cos(\sqrt{5}) \right) 
    + i (\sigma^x_1 \sigma^x_2 + \sigma^y_1 \sigma^y_2)  \frac{\sin(\sqrt{5})}{\sqrt{5}}.
\end{multline}
The trace is therefore
\begin{align}
    \Tr{e^{i\Gamma}} = 2\left( \cos(1) + \cos(\sqrt{5}) \right),
\end{align}
which is real, so $\vert \Tr{e^{i\Gamma}} \vert = \Tr{e^{i\Gamma}}$ and we now only need to maximise the imaginary part of $\Tr{e^{i\Gamma} \textbf{K}}$ to maximise the expression in Eq.~\eqref{eq:max_f1}.
We introduce the notation $\bm{\sigma}_{(2)}$ as a vector of all combinations of the tensor product of two Pauli matrices (including the product of two identities), and $\hat{k}$ is a unit vector with the same dimension of 16, such that $\textbf{K} = \hat{k} \cdot \bm{\sigma}_{(2)}$. The elements of $\hat{k}$ are labelled as $\hat{k}_{\alpha, \beta}$, where $\alpha,\beta = 0,1,2,3$ and refer to $I, \sigma^{x}, \sigma^{y}, \sigma^{z}$ respectively. Noting that only the trace over resulting identity terms gives a non-zero contribution, we find
\begin{multline}
    \label{eq:trace_expression}
    \textrm{Im}\left[ \Tr{e^{i\Gamma} \hat{k} \cdot \bm{\sigma}_{(2)} } \right] = 2 \hat{k}_{3,0} \left( \sin(1) - \frac{\sin(\sqrt{5})}{\sqrt{5}}\right)+ 2 \hat{k}_{0,3} \left( \sin(1) + \frac{\sin(\sqrt{5})}{\sqrt{5}}\right) + 4 \hat{k}_{1,1}  \frac{\sin(\sqrt{5})}{\sqrt{5}}\\
    + 4 \hat{k}_{2,2} \frac{\sin(\sqrt{5})}{\sqrt{5}}. 
\end{multline}
The expression can be written in the form
\begin{align}
    Q(\hat{k}) = \bm{a}\cdot\hat{k}, \quad \Vert \hat{k} \Vert = 1,
\end{align}
which has a maximum value 
\begin{align}
    \max_{\Vert\hat{k}\Vert=1} Q(\hat{k}) = \Vert \bm{a} \Vert,
\end{align}
at 
\begin{align}
    \hat{k}_\textrm{max} = \frac{\bm{a}}{\Vert \bm{a}\Vert}. 
\end{align}
The trace expression of Eq.~\eqref{eq:trace_expression} therefore has a maximum value at $\hat{k}_{3,0} = 0.30054$, $\hat{k}_{0,3} = 0.73248$, $\hat{k}_{1,1} = 0.43194$, $\hat{k}_{2,2} = 0.43194$,
giving an overlap with the geodesic vector in the tangent space of 
\begin{align}
    \hat{k} \cdot \frac{\bm{\gamma} }{\Vert \bm{\gamma} \Vert} = 0.922 < 1.
\end{align}
This example shows that the update step in GRAPE by maximising the fidelity locally does not follow the geodesic direction to the final solution. In this example, we assume Hamiltonian terms with no restrictions. We typically find larger deviations of the fidelity update direction from the geodesic direction for complicated forms of $\Gamma$. This provides evidence as to why GEOPE works more effectively than GRAPE, with significant improvements in convergence times.

\clearpage
\section{\label{sec:grape_methods}GRAPE methods}
Gradient ascent pulse engineering (GRAPE) is an algorithm that was originally introduced in the context of pulse design for NMR spectroscopy~\cite{khaneja_optimal_2005}, but has found wider applications~\cite{koch2022}. It is still considered the state-of-the-art~\cite{ansel_introduction_2024}, which is why we choose to compare our method to it.

\subsection{\label{sec:adam}Adam}
The most basic version of GRAPE uses standard gradient ascent to minimise the infidelity $I(\bPhi, V)$. This is achieved by calculating the gradient of the infidelity,
\begin{align}
    I(\bPhi,V) = 1 - F(\bPhi, V),\quad \boldsymbol{g}_{l,k} =\partial_{x_{l,k}} I(\bTheta, V).
\end{align}
Note that here, we work directly in the restricted space $\mathcal{H}$, as indicated by the parameterisation in $\bPhi$. To escape local minima, we make use of the stochastic gradient descent variant Adam~\cite{kingma_adam_2017}, which adaptively scales the gradients. The default hyperparameters of $\beta_1=0.9$ and $\beta_2=0.999$ rarely require tuning in practice hence we leave these unchanged. We present the method used in our simulations in Algorithm~\ref{alg:adam}.

\begin{algorithm}
\caption{GRAPE (Adam).}\label{alg:adam}
\KwIn{$V$, $\bm{\Phi}, R_{\mathcal{H}},  \varepsilon, \lambda, M$}
\KwOut{$\bm{\Phi}$}

\text{Define evolution:} $U_G(\bPhi) = U(\bPhi_K)...U(\bPhi_1)$

\text{Define an infidelity function:} $I(\bPhi, V) = 1 - \Tr{U_G(\bPhi)^\dagger V}/N$

\text{Initialise Adam moments per layer:} $\boldsymbol{\mu}_{l,k}=0$, $\boldsymbol{v}_{l,k}=0$\\
$\beta_1\gets 0.9$\\
$\beta_2\gets 0.999$\\

\For{$m \in\{1,2\ldots, M\}$}{
    \text{Compute the gradient at $\bPhi$:}\\
    {
    $\boldsymbol{g}_{l,k} \gets \partial_{x_{l,k}} I(\bPhi, V)$
    }

    \text{Calculate Adam moments:}\\
    
    $\boldsymbol{\mu}_{l,k}\gets \beta_1\cdot \boldsymbol{\mu}_{l,k} + (1-\beta_1)\cdot \boldsymbol{g}_{l,k}$\\
    $\boldsymbol{v}_{l,k}\gets \beta_2\cdot \boldsymbol{v}_{l,k} + (1-\beta_2)\cdot \boldsymbol{g}_{l,k}^2$\\
    $\hat{\boldsymbol{\mu}}_{l,k}\gets \boldsymbol{\mu}_{l,k} / (1 - \beta_1^t)$\\
    $\hat{\boldsymbol{v}}_{l,k}\gets \boldsymbol{v}_{l,k} / (1 - \beta_2^t)$\\

    \text{Update parameters:}\\
    $\bPhi \gets \bPhi - \lambda \cdot  \hat{\boldsymbol{\mu}}/(\sqrt{\hat{\boldsymbol{v}}} + \mathrm{eps})$
        
    \If{$I(\bPhi, V) <\varepsilon$}{
    break
    }
}
\end{algorithm}
\clearpage
\subsection{\label{sec:nr}Newton-Raphson}
To speed up the convergence of the gradient descent method, we can make use of a second order approach. In the Newton-Rhapson method, the Hessian is used to include curvature information of the cost function in the optimisation. If we define the Hessian as
\begin{align}
    \mathbf{H}_{l,k, l',k'}= \partial_{x_{l',k'}} \partial_{x_{l,k}}I(\bPhi,V),
\end{align}
then the Newton-Raphson update obtained by solving the linear system
\begin{align}
    \mathbf{H}\boldsymbol{u} = \boldsymbol{g},\label{eq:newt}
\end{align}
which gives the parameter update
\begin{align}
    \bPhi \gets \bPhi - \lambda\boldsymbol{u}\label{eq:hess_update}
\end{align}
for some chosen step size $\lambda$. Note that the Hessian can be calculated efficiently for our control problem with a similar method as in ~\label{eq:aux_jac} (see \cite{goodwin_auxiliary_2015}). 

Note that the update of Eq.~\eqref{eq:hess_update} will only result in a decrease of the infidelity if $H$ is positive-definite, which cannot be guaranteed for our problem. The most straightforward regularisation approach involves shifting the spectrum of $\mathbf{H}$ by a constant $\delta$~\cite{goodwin_modified_2016}, which is what we do here.
A potential complication with this approach is that if $\delta$ is too large and the Hessian is ill-conditioned, we suppress the curvature of the problem which results in slow convergence. 
To ensure that we make the most out of each computation of the Hessian, we use a backtracking line-search with Armijo criterion to find the optimal step size $\lambda$ at each iteration of the algorithm~\cite{armijo1966minimization}.

\begin{algorithm}
\caption{GRAPE.}\label{alg:nr}
\KwIn{$V$, $\bm{\Phi}, R_{\mathcal{H}},  \varepsilon, \delta, M$}
\KwOut{$\bm{\Phi}$}

\text{Define evolution:} $U_G(\bPhi) = U(\bPhi_K)...U(\bPhi_1)$

\text{Define an infidelity function:} $I(\bPhi, V) = 1- \Tr{U_G(\bPhi)^\dagger V}/N$

\For{$t \in\{1,2\ldots, M\}$}{
    \text{Compute the gradient and Hessian at $\bPhi$:}\\
    $\boldsymbol{g}_{l,k} \gets \partial_{x_{l,k}} I(\bPhi,)$\\
    $ \mathbf{H}_{l,k, l',k'}\gets \partial_{x_{l',k'}} \partial_{x_{l,k}}I(\bPhi,V)$

    \text{Diagonalise Hessian:}\\
    $\boldsymbol{\Sigma}, \boldsymbol{Q}\gets \texttt{eigh}(\mathbf{H})$\\
    \text{Regularise spectrum}\\
    $\boldsymbol{\sigma} \gets \max(\text{eps}, \delta - \min(\boldsymbol{\Sigma}))$\\
    $\boldsymbol{\Sigma} \gets \boldsymbol{\Sigma}+\boldsymbol{\sigma}$\\
    \text{Cholesky solve from lower triangular matrix}\\
    $\boldsymbol{L} \gets \texttt{cholesky}(\boldsymbol{Q} \boldsymbol{\Sigma} \boldsymbol{Q}^\dagger)$\\
    $\boldsymbol{u} \gets \texttt{solve}(L, \boldsymbol{g})$\\
    \text{Update parameters with backtracking linesearch:}\\
    \FuncSty{Minimise}$\left(I(\bPhi + \lambda\boldsymbol{u}), \lambda\right)$\\    
    \If{$I(\bPhi, V) < \varepsilon$}{
    break
    }
}
\end{algorithm}
\clearpage
\subsection{\label{sec:rfo}Rational Function Optimization (RFO)}
In the Newton-Raphson method, we approximate the function of interest with a Taylor expansion and derive an optimal step criterion based on the second-order expansion. Instead, we can consider a Rational Function Approximation (RFO)~\cite{banerjee1985search} to derive an update state, which requires constructing the auxiliary Hessian~\cite{goodwin_modified_2016}
\begin{align}
    \mathbf{H}_{\mathrm{aug}}=
    \left(\begin{array}{cc}
        \alpha^2 \mathbf{H} & \alpha \boldsymbol{g}\\
        \alpha \boldsymbol{g}^T & 0
    \end{array}\right),
\end{align}
which we then proceed to regularise as before to obtain a matrix $\mathbf{H}_{\mathrm{aug}}^{\mathrm{reg}} / \alpha^2$. We then take the top left block of this regularised matrix, denoted by $\mathbf{H}^{\mathrm{reg}}$ and calculate the update step with Eq.~\eqref{eq:newt}. 

The advantage of the RFO approach is that in addition to ensuring the positive-definiteness of the Hessian, the parameter $\alpha$ acts as a damping factor that can be decreased to lower the condition number of $\mathbf{H}^{\mathrm{reg}}$. In practice, we use a subroutine at each step to lower the condition number adaptively. In particular, we take
\begin{align}
    \alpha_{i+1}=\phi \alpha_i 
\end{align}
with $\alpha=1$ and $\phi=0.9$. We then set a target condition number of $\kappa$, which is the only hyperparameter of this approach. We fix the maximum number of iterations for this subroutine to $300$, which we found to work well in practice for the problems studied here.
\begin{algorithm}
\caption{GRAPE (RFO).}\label{alg:rfo}
\KwIn{$V$, $\bm{\Phi}, R_{\mathcal{H}},  \varepsilon, \delta, M$}
\KwOut{$\bm{\Phi}$}

\text{Define evolution:} $U_G(\bPhi) = U(\bPhi_K)...U(\bPhi_1)$

\text{Define an infidelity function:} $I(\bPhi, V) = 1- \Tr{U_G(\bPhi)^\dagger V}/N$\\
$\alpha\gets 1$\\
\For{$t \in\{1,2\ldots, M\}$}{
    \text{Compute the gradient and Hessian at $\bPhi$:}\\
    $\boldsymbol{g}_{l,k} \gets \partial_{x_{l,k}} I(\bPhi,)$\\
    $ \mathbf{H}_{l,k, l',k'}\gets \partial_{x_{l',k'}} \partial_{x_{l,k}}I(\bPhi,V)$\\
    \text{Condition augmented Hessian:}\\
    \For{$i \in\{1,2\ldots, 300\}$}{
        $
        \mathbf{H}_{\mathrm{aug}}\gets
        \left(\begin{array}{cc}
            \alpha^2 \mathbf{H} & \alpha \boldsymbol{g}\\
            \alpha \boldsymbol{g}^T & 0
        \end{array}\right)$\\
        \text{Diagonalise augmented Hessian:}\\
        $\boldsymbol{\Sigma}, \boldsymbol{Q}\gets \texttt{eigh}(\mathbf{H}_{\mathrm{aug}})$\\
        \text{Regularise spectrum}\\
        $\boldsymbol{\sigma} \gets \max(\text{eps}, \min(\boldsymbol{\Sigma}))$\\
        $\boldsymbol{\Sigma} \gets \boldsymbol{\Sigma}+\boldsymbol{\sigma}$\\
        $\mathbf{H}^{\mathrm{reg}}_{\mathrm{aug}}\gets \boldsymbol{Q} \boldsymbol{\Sigma} \boldsymbol{Q}^\dagger$\\
        \text{Get top left block of augmented Hessian}\\
        $\mathbf{H}^{\mathrm{reg}}\gets (\mathbf{H}^{\mathrm{reg}}_{\mathrm{aug}})_{11}/\alpha^2$\\
        $\alpha\gets\phi\alpha$\\
        \If{$\norm{(\mathbf{H}^{\mathrm{reg}})^{-1}}\norm{\mathbf{H}^{\mathrm{reg}}}<\kappa$}{
        break
        }
    }
    \text{Cholesky solve from regularised Hessian}\\
    $\boldsymbol{L} \gets \texttt{cholesky}(\mathbf{H}^{\mathrm{reg}})$\\
    $\boldsymbol{u} \gets \texttt{solve}(L, \boldsymbol{g})$\\
    \text{Update parameters with backtracking linesearch:}\\
    \FuncSty{Minimise}$\left(I(\bPhi + \lambda\boldsymbol{u}), \lambda\right)$\\    
    \If{$I(\bPhi, V) < \varepsilon$}{
    break
    }
}
\end{algorithm}

\section{\label{sec:complexity}Computational complexity}
Computing the fidelity for both GRAPE and GEOPE has a computational cost of $O(LN^3)$, which is also the case for computing $U_\textrm{G}(\bPhi)$, where $L$ is the number of piecewise layers. GEOPE calculates and stores the Jacobian for each parameter, while GRAPE calculates the gradient of the fidelity function for each parameter. 
% Both are implemented with automatic differentiation. 
The computational complexity cost of GEOPE is $O(KLN^4)$, with $K=\vert\mathcal{H}\vert$ being the number of parameters in the restriction set, and $KL$ being the total number of control parameters. The additional factor of $N$ is due to the number of rows in the Jacobian, the $N^3$ being the cost of computing the evolution. The cost of computing the gradient for GRAPE is $O(KLN^3)$, which is the same order as the cost of the fidelity function. GEOPE has to also compute the geodesic, which is simply the matrix logarithm, which is also $O(N^3)$. The update step for GRAPE is matrix multiplication giving $O(KLN^3)$. GEOPE solves the linear least-squares convex optimisation problem, which also has complexity $O(KLN^4)$~\cite[Section~1.2.2]{boyd2004convex}. Overall, both complexities are polynomial in $N$ with similar orders, GRAPE is $O(KLN^3)$ and GEOPE is $O(KLN^4)$. 

\section{\label{sec:hyperparameters}Hyperparameters}
Each pulse engineering method contains a hyperparameter. In the case of Newton-Raphson TRM, Newton-Raphson RFO, and GEOPE, reasonable choices in our implementations have distilled the number of hyperparameters to just one. The hyperparameters for GRAPE are described in App.~\ref{sec:grape_methods}. The choice of hyperparameters for the second order Newton-Raphson methods can have a large effect on the behaviour of the method, see Figs.~\ref{fig:grape_vs_geope_parameters} and~\ref{fig:grape_vs_geope_parameters_ccx} in the main text. The hyperparameter for GEOPE is the maximum step size $s$. However, as we shall see, changing $s$ from the optimal value does not have a significant impact on the efficacy of GEOPE to find a solution.

\subsection{\label{sec:bayesian_optimization}Bayesian optimisation}
The hyperparameters are found using Bayesian optimisation~\cite{frazier_tutorial_2018}. Bayesian optimisation is often used for problems where the cost function is a derivative-free black box and evaluating the cost is expensive. Bayesian optimisation uses a distribution of functions to fit the data with uncertainty and then an acquisition function to determine the best next parameter guess. The aim of Bayesian optimisation is to discover the optimal (maximum or minimum) of the black box function in the minimum number of evaluations. The number of evaluations is minimised by balancing exploration of the parameter space with exploitation of the current optimum. 

Bayesian optimisation places a Gaussian process prior on the black box function. A number of initial points, in our case $n_0 = 5$, are randomly chosen within the selected bounds on the hyperparameter. The Guassian process gives a distribution for the value of the black box as a function of the parameters, with a mean and standard deviation, which allows a $95\%$ confidence interval to be calculated. After each evaluation of the black box function, the Gaussian process distribution is updated, typically reducing the uncertainty in the measured region of the parameter space. A parameter $\alpha_{\textrm{BO}}$ reflects the level of noise inherent in the data of the Gaussian process prior.
The selection of the next point uses an acquisition function, which in our case is the \emph{upper confidence bound} method~\cite{srinivas_gaussian_2010}. The value of the upper confidence bound parameter $\kappa_{\textrm{BO}}$ determines the balance between exploration of the parameter space and exploitation of the current optimal value. Essentially $\kappa_{\textrm{BO}}$ will influence how evenly distributed the observations are over the full hyperparameter space.

First, we have to decide the figure of merit to optimise for the hyperparameter choice. Generally, for 3-qubit gates, we have found that GRAPE with Adam takes about 200 iterations to find a solution when it works well, see Figs.~\ref{fig:grape_vs_geope_parameters} and~\ref{fig:grape_vs_geope_parameters_ccx}. As before, we define finding a \emph{solution} as the infidelity reaching $\varepsilon = 10^{-9}$. We therefore determine that in the case that the Newton-Raphson methods find solutions, they should improve on finding a solution in 200 iterations. We want to include both whether a solution is found and the number of iterations for a solution to be found. Taking a cumulative infidelity up to a maximum of 200 iterations for each sample balances these considerations, 
\begin{align}
    \label{eq:mean_cumulative_infidelity}
    C(p) = \frac{1}{N_a}\sum_{a=1}^{N_a} \sum_{m=1}^{M_{a}} I(\bPhi_a^{(m)}, V),
\end{align}
where $p$ is the hyperparameter, and $M_a$ is 200 if a solution is not reached, otherwise it is the number of iterations to reach the solution for that particular sample $a$. Each \emph{observation} is $N_a$ samples. $I(\bPhi_a^{(m)}, V)$ is the infidelity as previously defined. A lower $C(p)$ means a better choice of $p$. This also provides a way to compare optimal control methods in terms of rate of convergence to solutions, a lower $\min_p\{C(p)\}$ indicates a better optimal control method. 

\begin{figure*}[htb!]
    \centering
    \subfloat[GRAPE: Adam]{\includegraphics[scale=0.64]{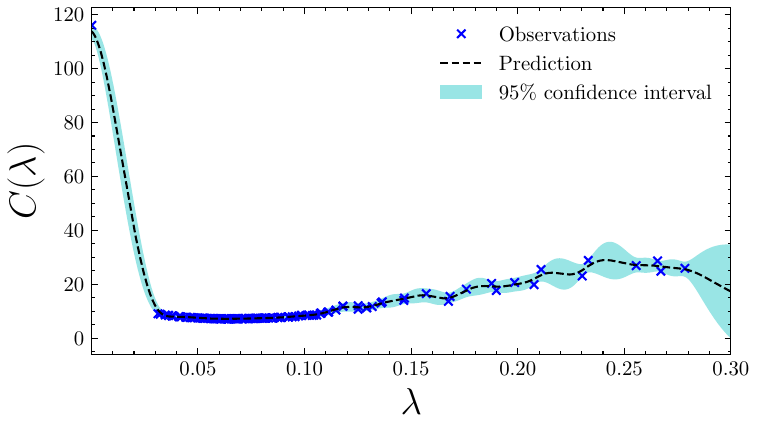}}\hspace{0.6cm}
    \subfloat[GRAPE: TRM]{\includegraphics[scale=0.64]{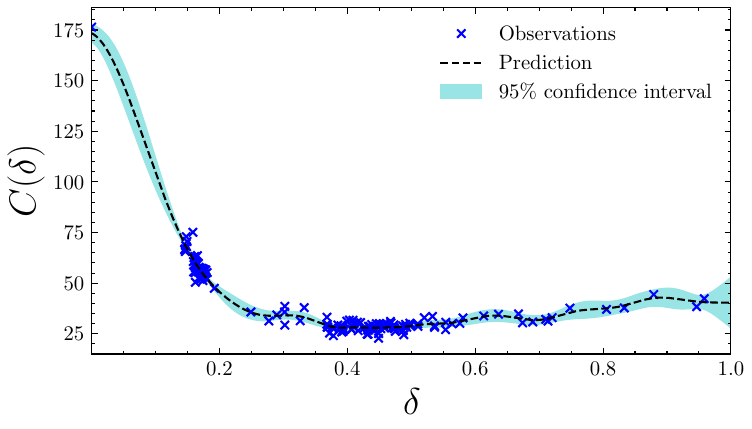}} \\
    \subfloat[GRAPE: RFO]{\includegraphics[scale=0.64]{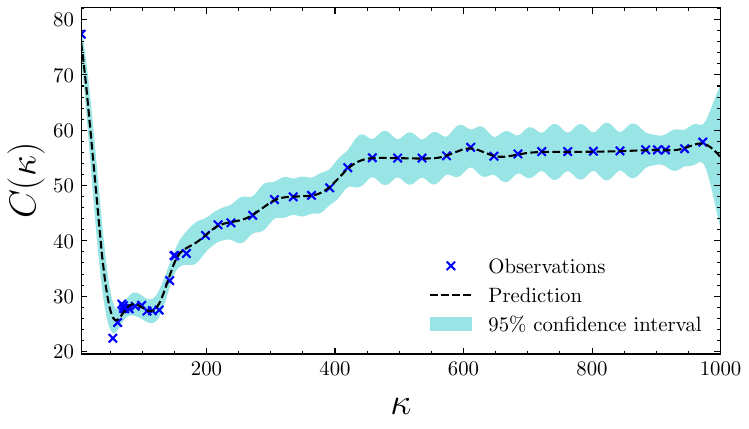}}\hspace{0.6cm}
    \subfloat[GEOPE]{\includegraphics[scale=0.64]{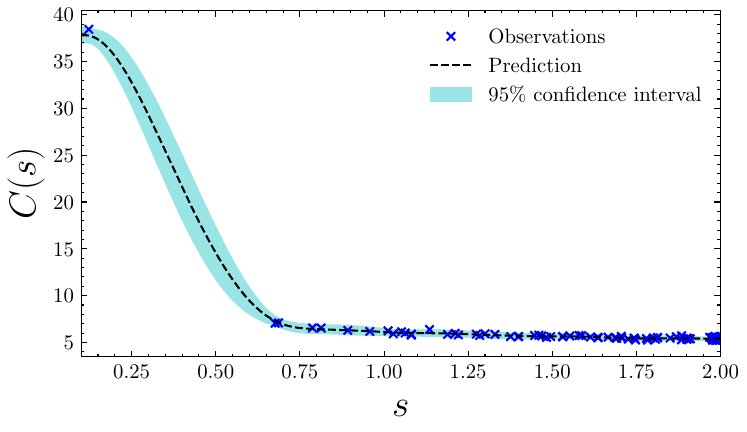}} 
    \caption{Hyperparameter search for a Toffoli gate with 12 piecewise steps ($L=12$) using GRAPE and GEOPE. Each observation is the mean cumulative infidelity of at least 50 samples with random $\bPhi^{(0)}$ parameter initialisations at various hyperparameter values with up to $M=200$ iterations, Eq.~\ref{eq:mean_cumulative_infidelity}. (a) GRAPE using Adam with different learning rates $\lambda$; (b) GRAPE using the Newton-Raphson TRM method with different $\delta$; (c) GRAPE using Newton-Raphson the RFO method with different condition numbers $\kappa$; (d) Our GEOPE algorithm with different maximum step size $\eta_\textrm{max}$. For GEOPE the upper bound for $\eta_\textrm{max}$ may constrain the absolute minimum. However, the solutions are found so quickly and The $95\%$ confidence interval and prediction come from the Bayesian optimisation method.}
    \label{fig:hyperparameter_ccx_12}
\end{figure*}

\begin{figure*}[htb!]
    \centering
    \subfloat[GRAPE: Adam]{\includegraphics[scale=0.64]{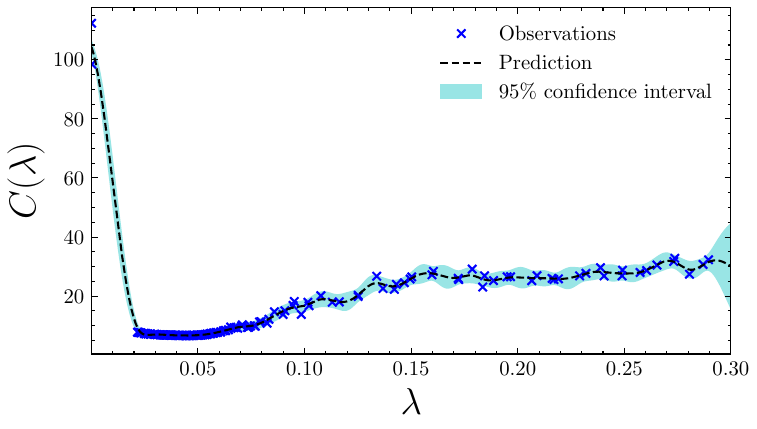}}\hspace{0.6cm}
    \subfloat[GRAPE: TRM]{\includegraphics[scale=0.64]{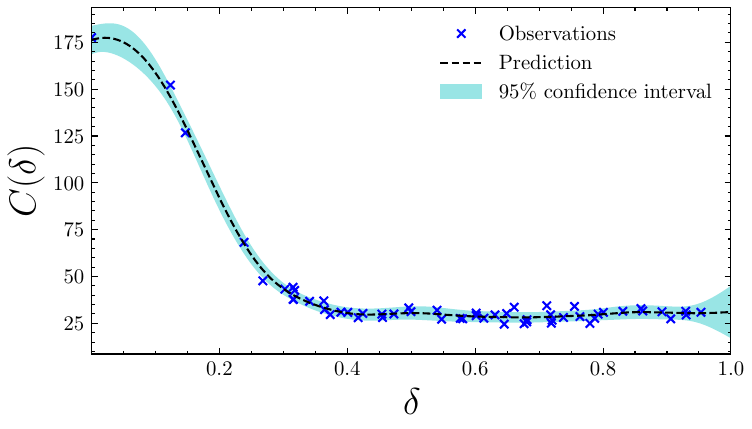}} \\
    \subfloat[GRAPE: RFO]{\includegraphics[scale=0.64]{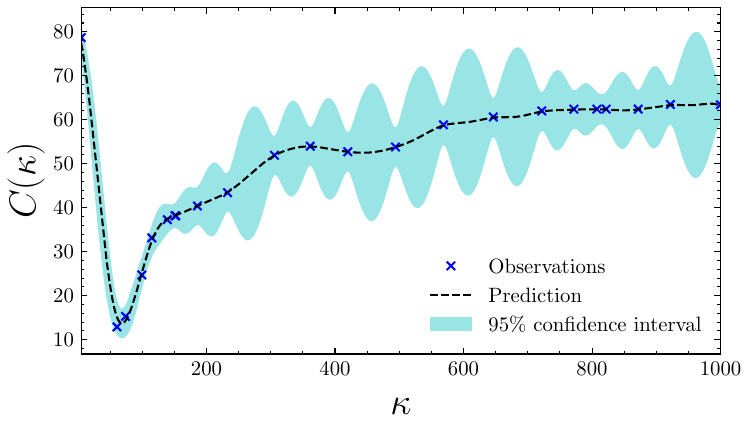}}\hspace{0.6cm}
    \subfloat[GEOPE]{\includegraphics[scale=0.64]{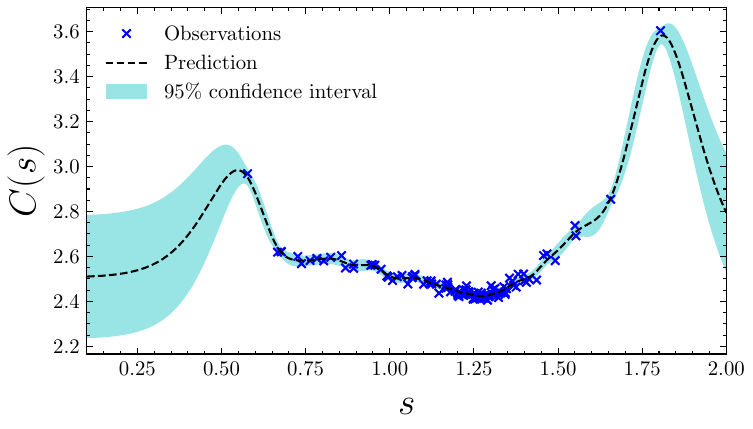}} 
    \caption{Hyperparameter search for a Toffoli gate with 20 piecewise steps ($L=20$) using GRAPE and GEOPE. Each observation is the mean cumulative infidelity of at least 50 samples with random $\bPhi^{(0)}$ parameter initialisations at a particular hyperparameter value with up to $M=200$ iterations, Eq.~\ref{eq:mean_cumulative_infidelity}. (a) GRAPE using Adam with different learning rates $\lambda$; (b) GRAPE using the Newton-Raphson TRM method with different $\delta$; (c) GRAPE using Newton-Raphson the RFO method with different condition numbers $\kappa$; (d) Our GEOPE algorithm with different maximum step size $\eta_\textrm{max}$. The $95\%$ confidence interval and prediction come from the Bayesian optimisation method.}
    \label{fig:hyperparameter_ccx_20}
\end{figure*}

The second choice is the values of the two important parameters relating to the Bayesian optimisation: the $\alpha_{\textrm{BO}}$ and $\kappa_{\textrm{BO}}$ values. After some experimentation, we found the values $\alpha_{\textrm{BO}}=0.02$ and $\kappa_{\textrm{BO}}=5$ led to successful hyperparameter searches. In Figs.~\ref{fig:hyperparameter_ccx_12} and~\ref{fig:hyperparameter_ccx_20}, we show the result of Bayesian optimisation over the hyperparameters for each optimal control method for generating Toffoli gates with 12 and 20 piecewise steps respectively. 

In Table~\ref{table:hyperparameters}, we list the optimal hyperparameters we have found for the 3-qubit QFT and the CCZ gates as well as the Toffoli. In all cases, we also give the optimal value of the mean cumulative infidelity $C(p)$ of Eq.~\ref{eq:mean_cumulative_infidelity}. The result clearly shows how GEOPE is significantly more effective in finding solutions for these optimal control problems.
\begin{table}
\begingroup
\setlength{\tabcolsep}{7pt} % Default value: 6pt
\renewcommand{\arraystretch}{1.2} % Default value: 1
\begin{tabular}{|l|l|c|c|l|l|}
\hline
Method                       & Gate                     & Piecewise steps K & Number of observations & Hyperparameter & $\min_p C(p)$ \\ \hline\hline
\multirow{6}{*}{GRAPE: Adam} & \multirow{2}{*} {Toffoli} & 12                & 129 & $\gamma =0.064$    &  7.17    \\ \cline{3-6} 
                             &                          & 20                & 159 & $\gamma =0.046$      &  6.65    \\ \cline{2-6} 
                             & \multirow{2}{*}{CCZ}     & 12                & 151 & $\gamma =0.050$      &  4.82    \\ \cline{3-6} 
                             &                          & 20                & 137 & $\gamma =0.029$      & 4.42     \\ \cline{2-6} 
                             & \multirow{2}{*}{3-QFT}   & 12                & 135 & $\gamma =0.107$      & 9.32     \\ \cline{3-6} 
                             &                          & 20                & 104 & $\gamma =0.06$      &  8.49    \\ \hline\hline
\multirow{6}{*}{GRAPE: NR}  & \multirow{2}{*}{Toffoli} & 12                & 151 & $\delta =0.449$      &  22.8    \\ \cline{3-6} 
                             &                          & 20                & 59 & $\delta = 0.645$      &  24.6    \\ \cline{2-6} 
                             & \multirow{2}{*}{CCZ}     & 12                & 80 & $\delta =0.736 $      & 6.40    \\ \cline{3-6} 
                             &                          & 20                & 26 & $\delta =0.993 $      &  11.0    \\ \cline{2-6} 
                             & \multirow{2}{*}{3-QFT}   & 12                & 133 & $\delta =0.205$      &  28.4    \\ \cline{3-6} 
                             &                          & 20                & 98 & $\delta =0.321$      &  30.7    \\ \hline\hline
\multirow{6}{*}{GRAPE: RFO}  & \multirow{2}{*}{Toffoli} & 12                & 44 & $\kappa =54.3$      & 22.4      \\ \cline{3-6} 
                             &                          & 20                & 25 & $\kappa =60.7$      &  12.9    \\ \cline{2-6} 
                             & \multirow{2}{*}{CCZ}     & 12                & 34 & $\kappa =45.3$      &  11.7    \\ \cline{3-6} 
                             &                          & 20                & 13 & $\kappa = 79.4$      & 21.1     \\ \cline{2-6} 
                             & \multirow{2}{*}{3-QFT}   & 12                & 37 & $\kappa =260.5$      &  36.7    \\ \cline{3-6} 
                             &                          & 20                & 5 & $\kappa =188.3$      &  61.4    \\ \hline\hline
\multirow{6}{*}{GEOPE}       & \multirow{2}{*}{Toffoli} & 12                & 114 & $\eta_\textrm{max}=1.98$               &  5.23    \\ \cline{3-6} 
                             &                          & 20                & 104 & $\eta_\textrm{max}=1.29$               &  2.41    \\ \cline{2-6} 
                             & \multirow{2}{*}{CCZ}     & 12                & 80 &$\eta_\textrm{max}=1.80$               &   5.28   \\ \cline{3-6} 
                             &                          & 20                & 159 &$\eta_\textrm{max}=1.42$               &   2.32   \\ \cline{2-6} 
                             & \multirow{2}{*}{3-QFT}   & 12                & 25 &$\eta_\textrm{max}=2.00$               &  6.52    \\ \cline{3-6} 
                             &                          & 20                & 104 & $\eta_\textrm{max}=1.25$               &  2.83    \\ \hline
\end{tabular}
\endgroup
\caption{Hyperparameters for each method for the Toffoli gate, the control-control-Z gate (CCZ) and the 3-qubit quantum Fourier transform (3-QFT) gate. The hyperparameters of maximum step size $s$ and Gram-Schmidt step size $g$ for GEOPE have been included, however these hyperparameters do not have a significant affect on the success of the method. }
\label{table:hyperparameters}
\end{table}
\FloatBarrier

\section{\label{sec:numerical_results}Numerical results}
\begin{figure*}[htb!]
    \centering
    \subfloat[GRAPE: Adam]{\includegraphics[scale=0.61]{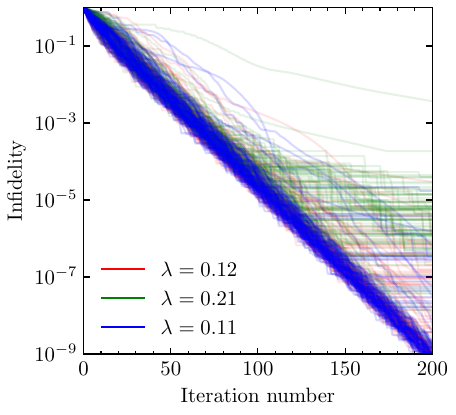}}
    \subfloat[GRAPE: Newton-Raphson]{\includegraphics[scale=0.61]{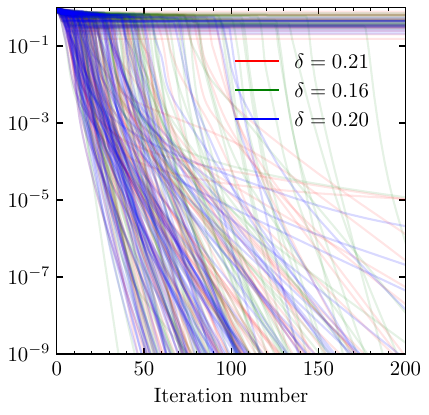}} 
    \subfloat[GRAPE: RFO]{\includegraphics[scale=0.61]{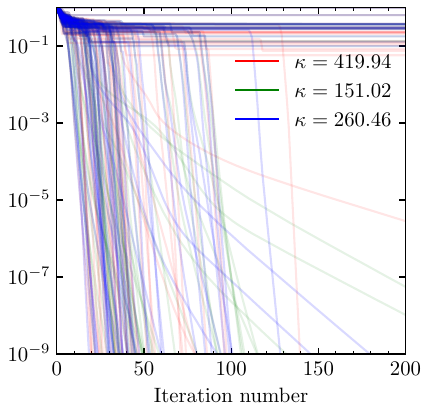}} 
    \subfloat[GEOPE]{\includegraphics[scale=0.61]{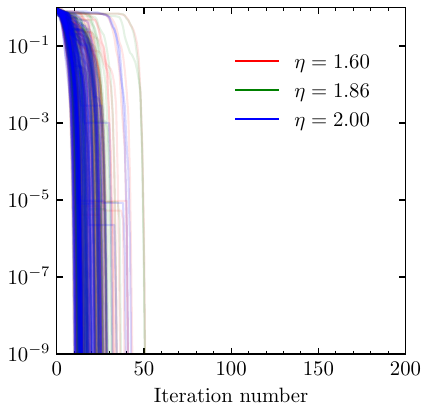}} 
    \caption{Designing a 3-qubit QFT gate with 12 piecewise steps ($L=12$) using GRAPE and GEOPE for the Rydberg model of Eq~\eqref{eq:rydberg_atom_array_model}. Infidelity at each iteration step is plotted for 100 samples at various hyperparameter values for: (a) GRAPE using Adam with different learning rates $\lambda$; (b) GRAPE using the Newton-Raphson TRM method with various $\delta$; (c) GRAPE using RFO method with various condition numbers $\kappa$; (d) Our GEOPE algorithm with various maximum step size $s$ and Gram-Schmidt step size $g$. Blue indicates the samples for the optimal hyperparameter value.}
    \label{fig:grape_vs_geope_parameters}
\end{figure*}

\begin{figure*}[htb!]
    \centering
    \subfloat[GRAPE: Adam]{\includegraphics[scale=0.61]{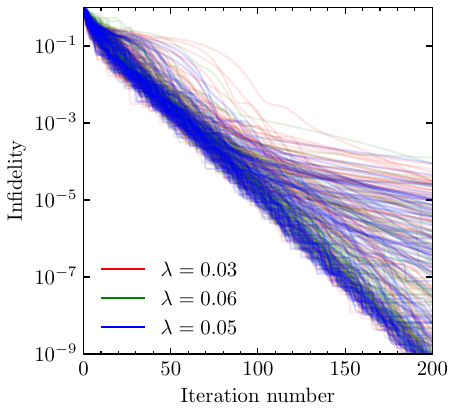}}
    \subfloat[GRAPE: Newton-Raphson]{\includegraphics[scale=0.61]{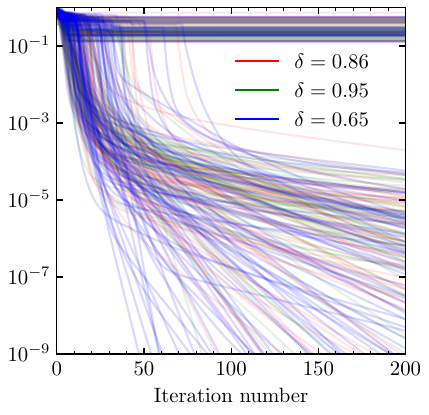}} 
    \subfloat[GRAPE: RFO]{\includegraphics[scale=0.61]{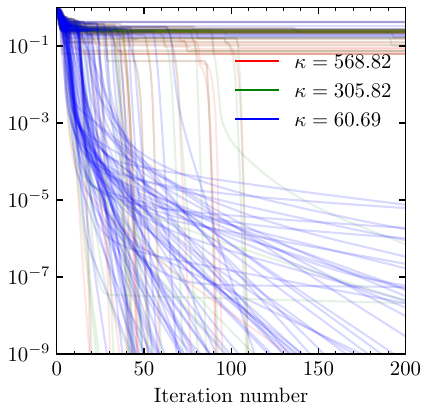}} 
    \subfloat[GEOPE]{\includegraphics[scale=0.61]{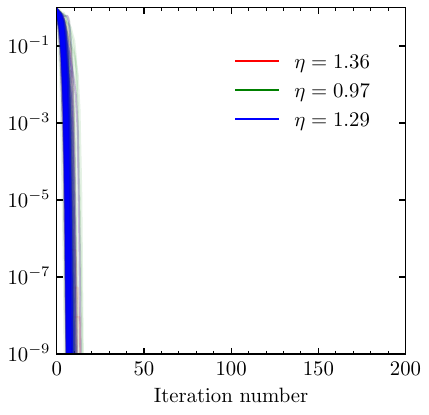}} 
    \caption{Designing a Toffoli gate with 20 piecewise steps ($L=20$) using GRAPE and GEOPE for the Rydberg model of Eq~\eqref{eq:rydberg_atom_array_model}. Infidelity at each iteration step is plotted for 100 samples at various hyperparameter values for: (a) GRAPE using Adam with different learning rates $\lambda$; (b) GRAPE using the Newton-Raphson method with various $\delta$; (c) GRAPE using RFO method with various condition numbers $\kappa$; (d) Our GEOPE algorithm with various maximum step size $s$ and Gram-Schmidt step size $g$. Blue indicates the samples for the optimal hyperparameter value.}
    \label{fig:grape_vs_geope_parameters_ccx}
\end{figure*}

In Figs.~\ref{fig:grape_vs_geope_parameters} and~\ref{fig:grape_vs_geope_parameters_ccx} samples for all the GRAPE methods and GEOPE are shown for finding 3-qubit QFT and Toffoli gate respectively. The optimal hyperparameter, as found using the method in App.~\ref{sec:bayesian_optimization}, is shown for each method with two other hyperparameter choices as a comparison. 

\section{\label{sec:hessian}Using the Hessian with GEOPE}
A change in the parameter vectors can be expanded to second order as
\begin{align}
     U_\textrm{G}(\bPhi^{(m)} + \bm{\delta}\bPhi^{(m)}) = U_\textrm{G}(\bPhi^{(m)}) + \sum_{k=1}^{K} \sum_{l \in \mathcal{H}} \mathbf{J}_{l,k} (\bPhi^{(m)}) \delta\phi^{(m)}_{l,k} + \frac{1}{2}  \sum_{r,l=1}^{K} \sum_{s,k \in \mathcal{H}}\delta\phi^{(m)}_{r,s}\mathbf{H}_{r, s, l,k} (\bPhi^{(m)}) \delta\phi^{(m)}_{l,k} + O(\delta\Phi^2),
\end{align}
where 
\begin{align}
    \mathbf{H}_{r,s,l,k}(\bm{\Phi}) = \frac{\partial^2 }{\partial x_{r,s} \partial x_{l,k}} \left( U(\bm{x}_K) ... U(\bm{x}_1) \right) \bigg\vert_{(\bm{x}_1, ..., \bm{x}_K) = \bPhi}
\end{align}
is the Hessian. The update objective therefore becomes
\begin{align}
\label{eq:update_condition_hessian}
    F_H(\bm{\delta\Phi}^{(m)}) = \min_{\bm{\delta\Phi}^{(m)}} \left\{ \bigg\Vert \sum_{k=1}^{K} \sum_{l \in \mathcal{H}} \mathbf{J}_{l,k} (\bPhi^{(m)}) \delta\phi^{(m)}_{l,k} + \frac{1}{2}  \sum_{r,k=1}^{K} \sum_{s,l \in \mathcal{H}}\delta\phi^{(m)}_{r,s}\mathbf{H}_{r, s, l,k} (\bPhi^{(m)}) \delta\phi^{(m)}_{l,k} - i U_\textrm{G}(\bPhi^{(m)}) \Gamma^{(m)} \bigg\Vert \right\}.
\end{align}
The rest of the GEOPE algorithm is unchanged. Computing the Hessian requires additional computational overhead for only marginal convergence improvements. 

\clearpage
\section{\label{sec:qft_solutions}QFT pulse sequences}
\begin{figure}[h]
    \centering
    \includegraphics[width=1\linewidth]{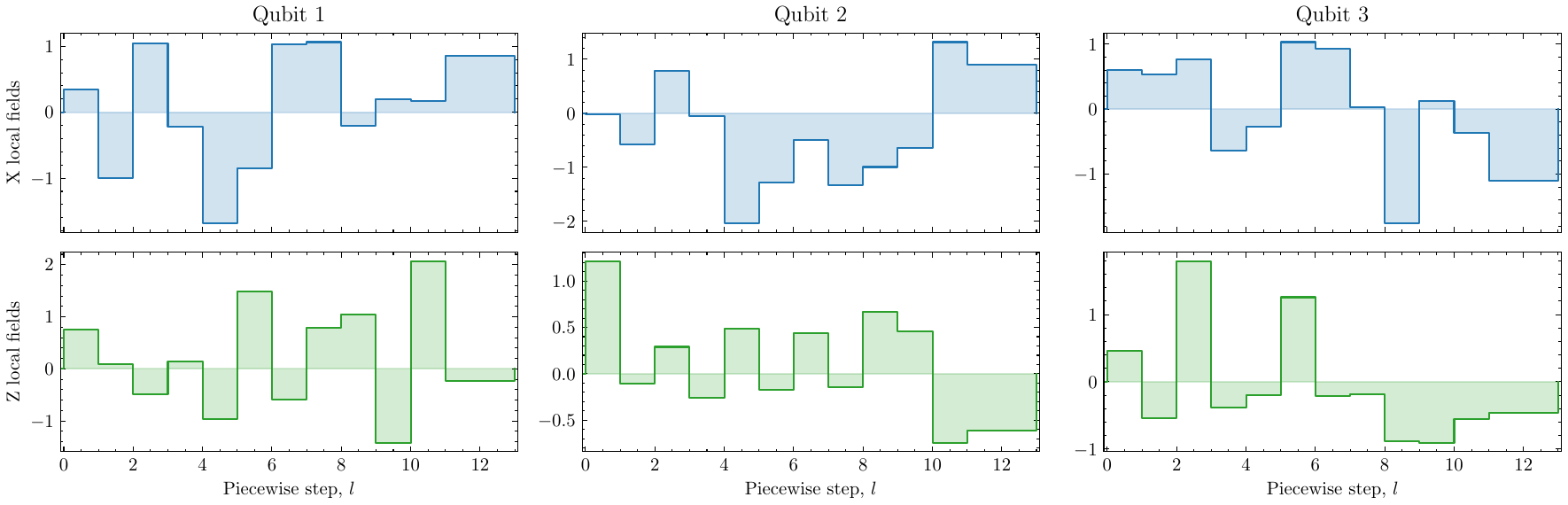}
    \caption{Solution for 3-qubit QFT gate with Rydberg model of Eq.~\eqref{eq:rydberg_atom_array_model} and qubit interaction given in Fig.~\ref{fig:rydberg_graphs}. GEOPE found this solution in 10 algorithmic steps. }
    \label{fig:qft3_geope_solution}
\end{figure}

\begin{figure}[h]
    \centering
    \includegraphics[width=1\linewidth]{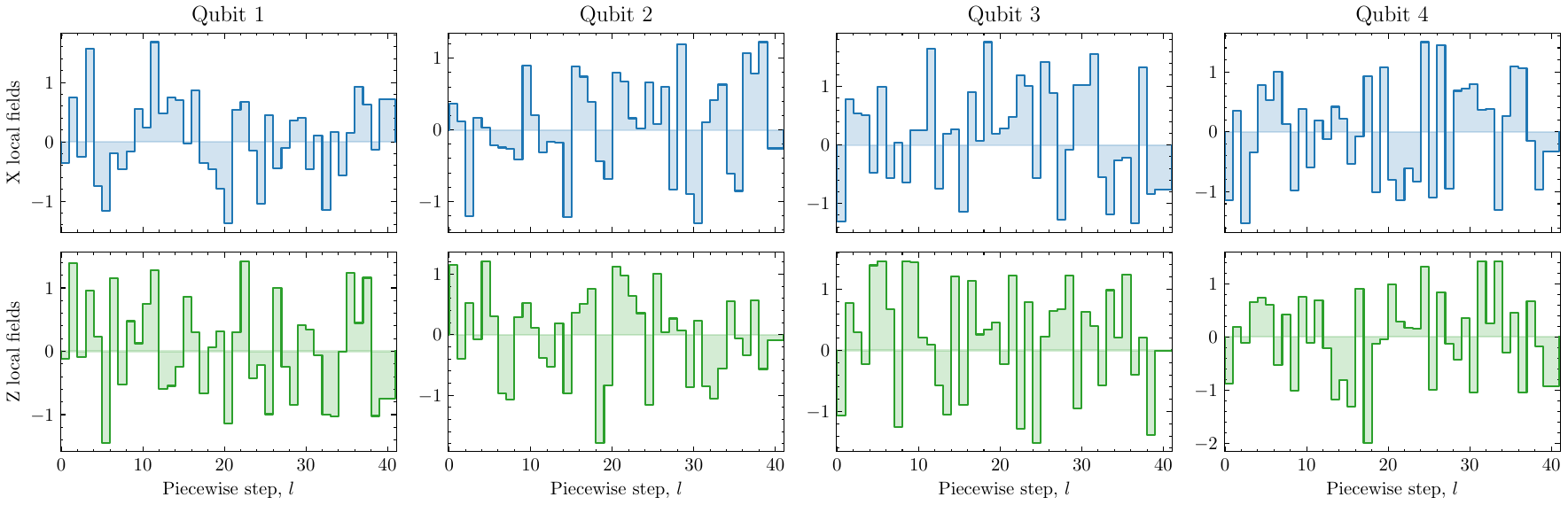}
    \caption{Solution for 4-qubit QFT gate with Rydberg model of Eq.~\eqref{eq:rydberg_atom_array_model} and qubit interaction given in Fig.~\ref{fig:rydberg_graphs}. GEOPE found this solution in 11 algorithmic steps.}
    \label{fig:qft4_geope_solution}
\end{figure}

The following plots show solutions for the QFT gate with Rydberg qubit interaction graphs given by Fig.~\ref{fig:rydberg_graphs} and with the Rydberg model of Eq.~\eqref{eq:rydberg_atom_array_model}. A solution for a 3- and 4-qubit QFT gate are given in Figs.~\ref{fig:qft3_geope_solution} and ~\ref{fig:qft4_geope_solution} respectively. The parameter values for QFT gates with 5 and 6 qubits are given in Ref.~\cite{our_data}.

\begin{thebibliography}{59}%
	\makeatletter
	\providecommand \@ifxundefined [1]{%
		\@ifx{#1\undefined}
	}%
	\providecommand \@ifnum [1]{%
		\ifnum #1\expandafter \@firstoftwo
		\else \expandafter \@secondoftwo
		\fi
	}%
	\providecommand \@ifx [1]{%
		\ifx #1\expandafter \@firstoftwo
		\else \expandafter \@secondoftwo
		\fi
	}%
	\providecommand \natexlab [1]{#1}%
	\providecommand \enquote  [1]{``#1''}%
	\providecommand \bibnamefont  [1]{#1}%
	\providecommand \bibfnamefont [1]{#1}%
	\providecommand \citenamefont [1]{#1}%
	\providecommand \href@noop [0]{\@secondoftwo}%
	\providecommand \href [0]{\begingroup \@sanitize@url \@href}%
	\providecommand \@href[1]{\@@startlink{#1}\@@href}%
	\providecommand \@@href[1]{\endgroup#1\@@endlink}%
	\providecommand \@sanitize@url [0]{\catcode `\\12\catcode `\$12\catcode
		`\&12\catcode `\#12\catcode `\^12\catcode `\_12\catcode `\%12\relax}%
	\providecommand \@@startlink[1]{}%
	\providecommand \@@endlink[0]{}%
	\providecommand \url  [0]{\begingroup\@sanitize@url \@url }%
	\providecommand \@url [1]{\endgroup\@href {#1}{\urlprefix }}%
	\providecommand \urlprefix  [0]{URL }%
	\providecommand \Eprint [0]{\href }%
	\providecommand \doibase [0]{https://doi.org/}%
	\providecommand \selectlanguage [0]{\@gobble}%
	\providecommand \bibinfo  [0]{\@secondoftwo}%
	\providecommand \bibfield  [0]{\@secondoftwo}%
	\providecommand \translation [1]{[#1]}%
	\providecommand \BibitemOpen [0]{}%
	\providecommand \bibitemStop [0]{}%
	\providecommand \bibitemNoStop [0]{.\EOS\space}%
	\providecommand \EOS [0]{\spacefactor3000\relax}%
	\providecommand \BibitemShut  [1]{\csname bibitem#1\endcsname}%
	\let\auto@bib@innerbib\@empty
	%</preamble>
	\bibitem [{\citenamefont {Goodwin}\ and\ \citenamefont
		{Vinding}(2023)}]{goodwin2023}%
	\BibitemOpen
	\bibfield  {author} {\bibinfo {author} {\bibfnamefont {D.~L.}\ \bibnamefont
			{Goodwin}}\ and\ \bibinfo {author} {\bibfnamefont {M.~S.}\ \bibnamefont
			{Vinding}},\ }\bibfield  {title} {\bibinfo {title} {Accelerated
			newton-raphson grape methods for optimal control},\ }\href
	{https://doi.org/10.1103/PhysRevResearch.5.L012042} {\bibfield  {journal}
		{\bibinfo  {journal} {Phys. Rev. Res.}\ }\textbf {\bibinfo {volume} {5}},\
		\bibinfo {pages} {L012042} (\bibinfo {year} {2023})}\BibitemShut {NoStop}%
	\bibitem [{\citenamefont {Kallies}(2018)}]{kallies2018}%
	\BibitemOpen
	\bibfield  {author} {\bibinfo {author} {\bibfnamefont {W.}~\bibnamefont
			{Kallies}},\ }{\selectlanguage {en}\emph {\bibinfo {title} {Concurrent
				optimization of robust refocused pulse sequences for magnetic resonance
				spectroscopy}}},\ \href {https://mediatum.ub.tum.de/1419323} {Ph.D. thesis},\
	\bibinfo  {school} {Technische Universität München} (\bibinfo {year}
	{2018})\BibitemShut {NoStop}%
	\bibitem [{\citenamefont {Werschnik}\ and\ \citenamefont
		{Gross}(2007)}]{werschnik_quantum_2007}%
	\BibitemOpen
	\bibfield  {author} {\bibinfo {author} {\bibfnamefont {J.}~\bibnamefont
			{Werschnik}}\ and\ \bibinfo {author} {\bibfnamefont {E.~K.~U.}\ \bibnamefont
			{Gross}},\ }\bibfield  {title} {{\selectlanguage {en}\bibinfo {title}
			{Quantum optimal control theory}},\ }\href
	{https://doi.org/10.1088/0953-4075/40/18/R01} {\bibfield  {journal} {\bibinfo
			{journal} {Journal of Physics B: Atomic, Molecular and Optical Physics}\
		}\textbf {\bibinfo {volume} {40}},\ \bibinfo {pages} {R175} (\bibinfo {year}
		{2007})}\BibitemShut {NoStop}%
	\bibitem [{\citenamefont {Glaser}\ \emph {et~al.}(2015)\citenamefont {Glaser},
		\citenamefont {Boscain}, \citenamefont {Calarco}, \citenamefont {Koch},
		\citenamefont {Köckenberger}, \citenamefont {Kosloff}, \citenamefont
		{Kuprov}, \citenamefont {Luy}, \citenamefont {Schirmer}, \citenamefont
		{Schulte-Herbrüggen}, \citenamefont {Sugny},\ and\ \citenamefont
		{Wilhelm}}]{glaser_training_2015}%
	\BibitemOpen
	\bibfield  {author} {\bibinfo {author} {\bibfnamefont {S.~J.}\ \bibnamefont
			{Glaser}}, \bibinfo {author} {\bibfnamefont {U.}~\bibnamefont {Boscain}},
		\bibinfo {author} {\bibfnamefont {T.}~\bibnamefont {Calarco}}, \bibinfo
		{author} {\bibfnamefont {C.~P.}\ \bibnamefont {Koch}}, \bibinfo {author}
		{\bibfnamefont {W.}~\bibnamefont {Köckenberger}}, \bibinfo {author}
		{\bibfnamefont {R.}~\bibnamefont {Kosloff}}, \bibinfo {author} {\bibfnamefont
			{I.}~\bibnamefont {Kuprov}}, \bibinfo {author} {\bibfnamefont
			{B.}~\bibnamefont {Luy}}, \bibinfo {author} {\bibfnamefont {S.}~\bibnamefont
			{Schirmer}}, \bibinfo {author} {\bibfnamefont {T.}~\bibnamefont
			{Schulte-Herbrüggen}}, \bibinfo {author} {\bibfnamefont {D.}~\bibnamefont
			{Sugny}},\ and\ \bibinfo {author} {\bibfnamefont {F.~K.}\ \bibnamefont
			{Wilhelm}},\ }\bibfield  {title} {{\selectlanguage {en}\bibinfo {title}
			{Training {Schrödinger}’s cat: quantum optimal control}},\ }\href
	{https://doi.org/10.1140/epjd/e2015-60464-1} {\bibfield  {journal} {\bibinfo
			{journal} {The European Physical Journal D}\ }\textbf {\bibinfo {volume}
			{69}},\ \bibinfo {pages} {279} (\bibinfo {year} {2015})}\BibitemShut
	{NoStop}%
	\bibitem [{\citenamefont {Boscain}\ \emph {et~al.}(2021)\citenamefont
		{Boscain}, \citenamefont {Sigalotti},\ and\ \citenamefont
		{Sugny}}]{boscain_introduction_2021}%
	\BibitemOpen
	\bibfield  {author} {\bibinfo {author} {\bibfnamefont {U.}~\bibnamefont
			{Boscain}}, \bibinfo {author} {\bibfnamefont {M.}~\bibnamefont {Sigalotti}},\
		and\ \bibinfo {author} {\bibfnamefont {D.}~\bibnamefont {Sugny}},\ }\bibfield
	{title} {\bibinfo {title} {Introduction to the {Pontryagin} {Maximum}
			{Principle} for {Quantum} {Optimal} {Control}},\ }\href
	{https://doi.org/10.1103/PRXQuantum.2.030203} {\bibfield  {journal} {\bibinfo
			{journal} {PRX Quantum}\ }\textbf {\bibinfo {volume} {2}},\ \bibinfo {pages}
		{030203} (\bibinfo {year} {2021})},\ \bibinfo {note} {publisher: American
		Physical Society}\BibitemShut {NoStop}%
	\bibitem [{\citenamefont {Ansel}\ \emph {et~al.}(2024)\citenamefont {Ansel},
		\citenamefont {Dionis}, \citenamefont {Arrouas}, \citenamefont {Peaudecerf},
		\citenamefont {Guérin}, \citenamefont {Guéry-Odelin},\ and\ \citenamefont
		{Sugny}}]{ansel_introduction_2024}%
	\BibitemOpen
	\bibfield  {author} {\bibinfo {author} {\bibfnamefont {Q.}~\bibnamefont
			{Ansel}}, \bibinfo {author} {\bibfnamefont {E.}~\bibnamefont {Dionis}},
		\bibinfo {author} {\bibfnamefont {F.}~\bibnamefont {Arrouas}}, \bibinfo
		{author} {\bibfnamefont {B.}~\bibnamefont {Peaudecerf}}, \bibinfo {author}
		{\bibfnamefont {S.}~\bibnamefont {Guérin}}, \bibinfo {author} {\bibfnamefont
			{D.}~\bibnamefont {Guéry-Odelin}},\ and\ \bibinfo {author} {\bibfnamefont
			{D.}~\bibnamefont {Sugny}},\ }\bibfield  {title} {\bibinfo {title}
		{Introduction to {Theoretical} and {Experimental} aspects of {Quantum}
			{Optimal} {Control}},\ }\href {https://doi.org/10.1088/1361-6455/ad46a5}
	{\bibfield  {journal} {\bibinfo  {journal} {Journal of Physics B: Atomic,
				Molecular and Optical Physics}\ }\textbf {\bibinfo {volume} {57}},\ \bibinfo
		{pages} {133001} (\bibinfo {year} {2024})},\ \bibinfo {note}
	{arXiv:2403.00532 [quant-ph]}\BibitemShut {NoStop}%
	\bibitem [{\citenamefont {Khaneja}\ \emph
		{et~al.}(2005{\natexlab{a}})\citenamefont {Khaneja}, \citenamefont {Reiss},
		\citenamefont {Kehlet}, \citenamefont {Schulte-Herbrüggen},\ and\
		\citenamefont {Glaser}}]{khaneja2005}%
	\BibitemOpen
	\bibfield  {author} {\bibinfo {author} {\bibfnamefont {N.}~\bibnamefont
			{Khaneja}}, \bibinfo {author} {\bibfnamefont {T.}~\bibnamefont {Reiss}},
		\bibinfo {author} {\bibfnamefont {C.}~\bibnamefont {Kehlet}}, \bibinfo
		{author} {\bibfnamefont {T.}~\bibnamefont {Schulte-Herbrüggen}},\ and\
		\bibinfo {author} {\bibfnamefont {S.~J.}\ \bibnamefont {Glaser}},\ }\bibfield
	{title} {\bibinfo {title} {Optimal control of coupled spin dynamics: design
			of nmr pulse sequences by gradient ascent algorithms},\ }\href
	{https://doi.org/https://doi.org/10.1016/j.jmr.2004.11.004} {\bibfield
		{journal} {\bibinfo  {journal} {Journal of Magnetic Resonance}\ }\textbf
		{\bibinfo {volume} {172}},\ \bibinfo {pages} {296} (\bibinfo {year}
		{2005}{\natexlab{a}})}\BibitemShut {NoStop}%
	\bibitem [{\citenamefont {Krotov}(1993)}]{krotov_global_1993}%
	\BibitemOpen
	\bibfield  {author} {\bibinfo {author} {\bibfnamefont {V.~F.}\ \bibnamefont
			{Krotov}},\ }\bibfield  {title} {{\selectlanguage {en}\bibinfo {title}
			{Global {Methods} in {Optimal} {Control} {Theory}}},\ }in\ \href
	{https://doi.org/10.1007/978-1-4612-0349-0_3} {{\selectlanguage {en}\emph
			{\bibinfo {booktitle} {Advances in {Nonlinear} {Dynamics} and {Control}: {A}
					{Report} from {Russia}}}}},\ \bibinfo {editor} {edited by\ \bibinfo {editor}
		{\bibfnamefont {A.~B.}\ \bibnamefont {Kurzhanski}}}\ (\bibinfo  {publisher}
	{Birkhäuser},\ \bibinfo {address} {Boston, MA},\ \bibinfo {year} {1993})\
	pp.\ \bibinfo {pages} {74--121}\BibitemShut {NoStop}%
	\bibitem [{\citenamefont {Palao}\ and\ \citenamefont
		{Kosloff}(2003)}]{palao_optimal_2003}%
	\BibitemOpen
	\bibfield  {author} {\bibinfo {author} {\bibfnamefont {J.~P.}\ \bibnamefont
			{Palao}}\ and\ \bibinfo {author} {\bibfnamefont {R.}~\bibnamefont
			{Kosloff}},\ }\bibfield  {title} {\bibinfo {title} {Optimal control theory
			for unitary transformations},\ }\href
	{https://doi.org/10.1103/PhysRevA.68.062308} {\bibfield  {journal} {\bibinfo
			{journal} {Physical Review A}\ }\textbf {\bibinfo {volume} {68}},\ \bibinfo
		{pages} {062308} (\bibinfo {year} {2003})},\ \bibinfo {note} {publisher:
		American Physical Society}\BibitemShut {NoStop}%
	\bibitem [{\citenamefont {Morzhin}\ and\ \citenamefont
		{Pechen}(2019)}]{morzhin_krotov_2019}%
	\BibitemOpen
	\bibfield  {author} {\bibinfo {author} {\bibfnamefont {O.}~\bibnamefont
			{Morzhin}}\ and\ \bibinfo {author} {\bibfnamefont {A.}~\bibnamefont
			{Pechen}},\ }\bibfield  {title} {\bibinfo {title} {Krotov {Method} for
			{Optimal} {Control} in {Closed} {Quantum} {Systems}},\ }\href
	{https://doi.org/10.1070/RM9835} {\bibfield  {journal} {\bibinfo  {journal}
			{Russian Mathematical Surveys}\ }\textbf {\bibinfo {volume} {74}},\ \bibinfo
		{pages} {851} (\bibinfo {year} {2019})},\ \bibinfo {note} {arXiv:1809.09562
		[quant-ph]}\BibitemShut {NoStop}%
	\bibitem [{\citenamefont {Caneva}\ \emph {et~al.}(2011)\citenamefont {Caneva},
		\citenamefont {Calarco},\ and\ \citenamefont
		{Montangero}}]{caneva_chopped_2011}%
	\BibitemOpen
	\bibfield  {author} {\bibinfo {author} {\bibfnamefont {T.}~\bibnamefont
			{Caneva}}, \bibinfo {author} {\bibfnamefont {T.}~\bibnamefont {Calarco}},\
		and\ \bibinfo {author} {\bibfnamefont {S.}~\bibnamefont {Montangero}},\
	}\bibfield  {title} {\bibinfo {title} {Chopped random-basis quantum
			optimization},\ }\href {https://doi.org/10.1103/PhysRevA.84.022326}
	{\bibfield  {journal} {\bibinfo  {journal} {Physical Review A}\ }\textbf
		{\bibinfo {volume} {84}},\ \bibinfo {pages} {022326} (\bibinfo {year}
		{2011})},\ \bibinfo {note} {publisher: American Physical Society}\BibitemShut
	{NoStop}%
	\bibitem [{\citenamefont {Müller}\ \emph {et~al.}(2022)\citenamefont
		{Müller}, \citenamefont {Said}, \citenamefont {Jelezko}, \citenamefont
		{Calarco},\ and\ \citenamefont {Montangero}}]{muller_one_2022}%
	\BibitemOpen
	\bibfield  {author} {\bibinfo {author} {\bibfnamefont {M.~M.}\ \bibnamefont
			{Müller}}, \bibinfo {author} {\bibfnamefont {R.~S.}\ \bibnamefont {Said}},
		\bibinfo {author} {\bibfnamefont {F.}~\bibnamefont {Jelezko}}, \bibinfo
		{author} {\bibfnamefont {T.}~\bibnamefont {Calarco}},\ and\ \bibinfo {author}
		{\bibfnamefont {S.}~\bibnamefont {Montangero}},\ }\bibfield  {title}
	{{\selectlanguage {en}\bibinfo {title} {One decade of quantum optimal control
				in the chopped random basis}},\ }\href
	{https://doi.org/10.1088/1361-6633/ac723c} {\bibfield  {journal} {\bibinfo
			{journal} {Reports on Progress in Physics}\ }\textbf {\bibinfo {volume}
			{85}},\ \bibinfo {pages} {076001} (\bibinfo {year} {2022})},\ \bibinfo {note}
	{publisher: IOP Publishing}\BibitemShut {NoStop}%
	\bibitem [{\citenamefont {Bukov}\ \emph {et~al.}(2018)\citenamefont {Bukov},
		\citenamefont {Day}, \citenamefont {Sels}, \citenamefont {Weinberg},
		\citenamefont {Polkovnikov},\ and\ \citenamefont
		{Mehta}}]{bukov_reinforcement_2018}%
	\BibitemOpen
	\bibfield  {author} {\bibinfo {author} {\bibfnamefont {M.}~\bibnamefont
			{Bukov}}, \bibinfo {author} {\bibfnamefont {A.~G.}\ \bibnamefont {Day}},
		\bibinfo {author} {\bibfnamefont {D.}~\bibnamefont {Sels}}, \bibinfo {author}
		{\bibfnamefont {P.}~\bibnamefont {Weinberg}}, \bibinfo {author}
		{\bibfnamefont {A.}~\bibnamefont {Polkovnikov}},\ and\ \bibinfo {author}
		{\bibfnamefont {P.}~\bibnamefont {Mehta}},\ }\bibfield  {title} {\bibinfo
		{title} {Reinforcement {Learning} in {Different} {Phases} of {Quantum}
			{Control}},\ }\href {https://doi.org/10.1103/PhysRevX.8.031086} {\bibfield
		{journal} {\bibinfo  {journal} {Physical Review X}\ }\textbf {\bibinfo
			{volume} {8}},\ \bibinfo {pages} {031086} (\bibinfo {year} {2018})},\
	\bibinfo {note} {publisher: American Physical Society}\BibitemShut {NoStop}%
	\bibitem [{\citenamefont {Day}\ \emph {et~al.}(2019)\citenamefont {Day},
		\citenamefont {Bukov}, \citenamefont {Weinberg}, \citenamefont {Mehta},\ and\
		\citenamefont {Sels}}]{day_glassy_2019}%
	\BibitemOpen
	\bibfield  {author} {\bibinfo {author} {\bibfnamefont {A.~G.}\ \bibnamefont
			{Day}}, \bibinfo {author} {\bibfnamefont {M.}~\bibnamefont {Bukov}}, \bibinfo
		{author} {\bibfnamefont {P.}~\bibnamefont {Weinberg}}, \bibinfo {author}
		{\bibfnamefont {P.}~\bibnamefont {Mehta}},\ and\ \bibinfo {author}
		{\bibfnamefont {D.}~\bibnamefont {Sels}},\ }\bibfield  {title} {\bibinfo
		{title} {Glassy {Phase} of {Optimal} {Quantum} {Control}},\ }\href
	{https://doi.org/10.1103/PhysRevLett.122.020601} {\bibfield  {journal}
		{\bibinfo  {journal} {Physical Review Letters}\ }\textbf {\bibinfo {volume}
			{122}},\ \bibinfo {pages} {020601} (\bibinfo {year} {2019})},\ \bibinfo
	{note} {publisher: American Physical Society}\BibitemShut {NoStop}%
	\bibitem [{\citenamefont {Khalid}\ \emph {et~al.}(2023)\citenamefont {Khalid},
		\citenamefont {Weidner}, \citenamefont {Jonckheere}, \citenamefont
		{Schirmer},\ and\ \citenamefont {Langbein}}]{khalid_sample-efficient_2023}%
	\BibitemOpen
	\bibfield  {author} {\bibinfo {author} {\bibfnamefont {I.}~\bibnamefont
			{Khalid}}, \bibinfo {author} {\bibfnamefont {C.~A.}\ \bibnamefont {Weidner}},
		\bibinfo {author} {\bibfnamefont {E.~A.}\ \bibnamefont {Jonckheere}},
		\bibinfo {author} {\bibfnamefont {S.~G.}\ \bibnamefont {Schirmer}},\ and\
		\bibinfo {author} {\bibfnamefont {F.~C.}\ \bibnamefont {Langbein}},\
	}\bibfield  {title} {\bibinfo {title} {Sample-efficient model-based
			reinforcement learning for quantum control},\ }\href
	{https://doi.org/10.1103/PhysRevResearch.5.043002} {\bibfield  {journal}
		{\bibinfo  {journal} {Physical Review Research}\ }\textbf {\bibinfo {volume}
			{5}},\ \bibinfo {pages} {043002} (\bibinfo {year} {2023})},\ \bibinfo {note}
	{publisher: American Physical Society}\BibitemShut {NoStop}%
	\bibitem [{\citenamefont {Mao}\ \emph {et~al.}(2023)\citenamefont {Mao},
		\citenamefont {Cheng}, \citenamefont {Xia}, \citenamefont {Oleś},\ and\
		\citenamefont {You}}]{mao_machine-learning-inspired_2023}%
	\BibitemOpen
	\bibfield  {author} {\bibinfo {author} {\bibfnamefont {M.-Y.}\ \bibnamefont
			{Mao}}, \bibinfo {author} {\bibfnamefont {Z.}~\bibnamefont {Cheng}}, \bibinfo
		{author} {\bibfnamefont {Y.}~\bibnamefont {Xia}}, \bibinfo {author}
		{\bibfnamefont {A.~M.}\ \bibnamefont {Oleś}},\ and\ \bibinfo {author}
		{\bibfnamefont {W.-L.}\ \bibnamefont {You}},\ }\bibfield  {title} {\bibinfo
		{title} {Machine-learning-inspired quantum optimal control of nonadiabatic
			geometric quantum computation via reverse engineering},\ }\href
	{https://doi.org/10.1103/PhysRevA.108.032616} {\bibfield  {journal} {\bibinfo
			{journal} {Physical Review A}\ }\textbf {\bibinfo {volume} {108}},\ \bibinfo
		{pages} {032616} (\bibinfo {year} {2023})},\ \bibinfo {note} {publisher:
		American Physical Society}\BibitemShut {NoStop}%
	\bibitem [{\citenamefont {Lin}\ \emph {et~al.}(2020)\citenamefont {Lin},
		\citenamefont {Sels}, \citenamefont {Ma},\ and\ \citenamefont
		{Wang}}]{lin2020stochastic}%
	\BibitemOpen
	\bibfield  {author} {\bibinfo {author} {\bibfnamefont {C.}~\bibnamefont
			{Lin}}, \bibinfo {author} {\bibfnamefont {D.}~\bibnamefont {Sels}}, \bibinfo
		{author} {\bibfnamefont {Y.}~\bibnamefont {Ma}},\ and\ \bibinfo {author}
		{\bibfnamefont {Y.}~\bibnamefont {Wang}},\ }\bibfield  {title} {\bibinfo
		{title} {Stochastic optimal control formalism for an open quantum system},\
	}\href@noop {} {\bibfield  {journal} {\bibinfo  {journal} {Physical Review
				A}\ }\textbf {\bibinfo {volume} {102}},\ \bibinfo {pages} {052605} (\bibinfo
		{year} {2020})}\BibitemShut {NoStop}%
	\bibitem [{\citenamefont {Villanueva}\ and\ \citenamefont
		{Kappen}(2024)}]{villanueva2024stochastic}%
	\BibitemOpen
	\bibfield  {author} {\bibinfo {author} {\bibfnamefont {A.}~\bibnamefont
			{Villanueva}}\ and\ \bibinfo {author} {\bibfnamefont {H.}~\bibnamefont
			{Kappen}},\ }\bibfield  {title} {\bibinfo {title} {Stochastic optimal control
			of open quantum systems},\ }\href@noop {} {\bibfield  {journal} {\bibinfo
			{journal} {arXiv preprint arXiv:2410.18635}\ } (\bibinfo {year}
		{2024})}\BibitemShut {NoStop}%
	\bibitem [{\citenamefont {Koch}\ \emph {et~al.}(2022)\citenamefont {Koch},
		\citenamefont {Boscain}, \citenamefont {Calarco}, \citenamefont {Dirr},
		\citenamefont {Filipp}, \citenamefont {Glaser}, \citenamefont {Kosloff},
		\citenamefont {Montangero}, \citenamefont {Schulte-Herbr{\"u}ggen},
		\citenamefont {Sugny},\ and\ \citenamefont {Wilhelm}}]{koch2022}%
	\BibitemOpen
	\bibfield  {author} {\bibinfo {author} {\bibfnamefont {C.~P.}\ \bibnamefont
			{Koch}}, \bibinfo {author} {\bibfnamefont {U.}~\bibnamefont {Boscain}},
		\bibinfo {author} {\bibfnamefont {T.}~\bibnamefont {Calarco}}, \bibinfo
		{author} {\bibfnamefont {G.}~\bibnamefont {Dirr}}, \bibinfo {author}
		{\bibfnamefont {S.}~\bibnamefont {Filipp}}, \bibinfo {author} {\bibfnamefont
			{S.~J.}\ \bibnamefont {Glaser}}, \bibinfo {author} {\bibfnamefont
			{R.}~\bibnamefont {Kosloff}}, \bibinfo {author} {\bibfnamefont
			{S.}~\bibnamefont {Montangero}}, \bibinfo {author} {\bibfnamefont
			{T.}~\bibnamefont {Schulte-Herbr{\"u}ggen}}, \bibinfo {author} {\bibfnamefont
			{D.}~\bibnamefont {Sugny}},\ and\ \bibinfo {author} {\bibfnamefont {F.~K.}\
			\bibnamefont {Wilhelm}},\ }\bibfield  {title} {\bibinfo {title} {Quantum
			optimal control in quantum technologies. strategic report on current status,
			visions and goals for research in europe},\ }\href
	{https://doi.org/10.1140/epjqt/s40507-022-00138-x} {\bibfield  {journal}
		{\bibinfo  {journal} {EPJ Quantum Technology}\ }\textbf {\bibinfo {volume}
			{9}},\ \bibinfo {pages} {19} (\bibinfo {year} {2022})}\BibitemShut {NoStop}%
	\bibitem [{\citenamefont {Genois}\ \emph {et~al.}(2024)\citenamefont {Genois},
		\citenamefont {Stevenson}, \citenamefont {Goss}, \citenamefont {Siddiqi},\
		and\ \citenamefont {Blais}}]{genois_quantum_2024}%
	\BibitemOpen
	\bibfield  {author} {\bibinfo {author} {\bibfnamefont {E.}~\bibnamefont
			{Genois}}, \bibinfo {author} {\bibfnamefont {N.~J.}\ \bibnamefont
			{Stevenson}}, \bibinfo {author} {\bibfnamefont {N.}~\bibnamefont {Goss}},
		\bibinfo {author} {\bibfnamefont {I.}~\bibnamefont {Siddiqi}},\ and\ \bibinfo
		{author} {\bibfnamefont {A.}~\bibnamefont {Blais}},\ }\href
	{https://doi.org/10.48550/arXiv.2410.22603} {\bibinfo {title} {Quantum
			optimal control of superconducting qubits based on machine-learning
			characterization}} (\bibinfo {year} {2024}),\ \bibinfo {note}
	{arXiv:2410.22603 [quant-ph]}\BibitemShut {NoStop}%
	\bibitem [{\citenamefont {Khaneja}\ \emph
		{et~al.}(2005{\natexlab{b}})\citenamefont {Khaneja}, \citenamefont {Reiss},
		\citenamefont {Kehlet}, \citenamefont {Schulte-Herbrüggen},\ and\
		\citenamefont {Glaser}}]{khaneja_optimal_2005}%
	\BibitemOpen
	\bibfield  {author} {\bibinfo {author} {\bibfnamefont {N.}~\bibnamefont
			{Khaneja}}, \bibinfo {author} {\bibfnamefont {T.}~\bibnamefont {Reiss}},
		\bibinfo {author} {\bibfnamefont {C.}~\bibnamefont {Kehlet}}, \bibinfo
		{author} {\bibfnamefont {T.}~\bibnamefont {Schulte-Herbrüggen}},\ and\
		\bibinfo {author} {\bibfnamefont {S.~J.}\ \bibnamefont {Glaser}},\ }\bibfield
	{title} {\bibinfo {title} {Optimal control of coupled spin dynamics: design
			of {NMR} pulse sequences by gradient ascent algorithms},\ }\href
	{https://doi.org/10.1016/j.jmr.2004.11.004} {\bibfield  {journal} {\bibinfo
			{journal} {Journal of Magnetic Resonance}\ }\textbf {\bibinfo {volume}
			{172}},\ \bibinfo {pages} {296} (\bibinfo {year}
		{2005}{\natexlab{b}})}\BibitemShut {NoStop}%
	\bibitem [{\citenamefont {Dolde}\ \emph {et~al.}(2014)\citenamefont {Dolde},
		\citenamefont {Bergholm}, \citenamefont {Wang}, \citenamefont {Jakobi},
		\citenamefont {Naydenov}, \citenamefont {Pezzagna}, \citenamefont {Meijer},
		\citenamefont {Jelezko}, \citenamefont {Neumann}, \citenamefont
		{Schulte-Herbrüggen}, \citenamefont {Biamonte},\ and\ \citenamefont
		{Wrachtrup}}]{dolde_high-fidelity_2014}%
	\BibitemOpen
	\bibfield  {author} {\bibinfo {author} {\bibfnamefont {F.}~\bibnamefont
			{Dolde}}, \bibinfo {author} {\bibfnamefont {V.}~\bibnamefont {Bergholm}},
		\bibinfo {author} {\bibfnamefont {Y.}~\bibnamefont {Wang}}, \bibinfo {author}
		{\bibfnamefont {I.}~\bibnamefont {Jakobi}}, \bibinfo {author} {\bibfnamefont
			{B.}~\bibnamefont {Naydenov}}, \bibinfo {author} {\bibfnamefont
			{S.}~\bibnamefont {Pezzagna}}, \bibinfo {author} {\bibfnamefont
			{J.}~\bibnamefont {Meijer}}, \bibinfo {author} {\bibfnamefont
			{F.}~\bibnamefont {Jelezko}}, \bibinfo {author} {\bibfnamefont
			{P.}~\bibnamefont {Neumann}}, \bibinfo {author} {\bibfnamefont
			{T.}~\bibnamefont {Schulte-Herbrüggen}}, \bibinfo {author} {\bibfnamefont
			{J.}~\bibnamefont {Biamonte}},\ and\ \bibinfo {author} {\bibfnamefont
			{J.}~\bibnamefont {Wrachtrup}},\ }\bibfield  {title} {{\selectlanguage
			{en}\bibinfo {title} {High-fidelity spin entanglement using optimal
				control}},\ }\href {https://doi.org/10.1038/ncomms4371} {\bibfield  {journal}
		{\bibinfo  {journal} {Nature Communications}\ }\textbf {\bibinfo {volume}
			{5}},\ \bibinfo {pages} {3371} (\bibinfo {year} {2014})},\ \bibinfo {note}
	{publisher: Nature Publishing Group}\BibitemShut {NoStop}%
	\bibitem [{\citenamefont {Gorman}\ \emph {et~al.}(2012)\citenamefont {Gorman},
		\citenamefont {Young},\ and\ \citenamefont
		{Whaley}}]{gorman_overcoming_2012}%
	\BibitemOpen
	\bibfield  {author} {\bibinfo {author} {\bibfnamefont {D.~J.}\ \bibnamefont
			{Gorman}}, \bibinfo {author} {\bibfnamefont {K.~C.}\ \bibnamefont {Young}},\
		and\ \bibinfo {author} {\bibfnamefont {K.~B.}\ \bibnamefont {Whaley}},\
	}\bibfield  {title} {\bibinfo {title} {Overcoming dephasing noise with robust
			optimal control},\ }\href {https://doi.org/10.1103/PhysRevA.86.012317}
	{\bibfield  {journal} {\bibinfo  {journal} {Physical Review A}\ }\textbf
		{\bibinfo {volume} {86}},\ \bibinfo {pages} {012317} (\bibinfo {year}
		{2012})},\ \bibinfo {note} {publisher: American Physical Society}\BibitemShut
	{NoStop}%
	\bibitem [{\citenamefont {Nielsen}\ \emph
		{et~al.}(2006{\natexlab{a}})\citenamefont {Nielsen}, \citenamefont {Dowling},
		\citenamefont {Gu},\ and\ \citenamefont {Doherty}}]{nielsen_quantum_2006}%
	\BibitemOpen
	\bibfield  {author} {\bibinfo {author} {\bibfnamefont {M.~A.}\ \bibnamefont
			{Nielsen}}, \bibinfo {author} {\bibfnamefont {M.~R.}\ \bibnamefont
			{Dowling}}, \bibinfo {author} {\bibfnamefont {M.}~\bibnamefont {Gu}},\ and\
		\bibinfo {author} {\bibfnamefont {A.~C.}\ \bibnamefont {Doherty}},\
	}\bibfield  {title} {\bibinfo {title} {Quantum {Computation} as {Geometry}},\
	}\href {https://doi.org/10.1126/science.1121541} {\bibfield  {journal}
		{\bibinfo  {journal} {Science}\ }\textbf {\bibinfo {volume} {311}},\ \bibinfo
		{pages} {1133} (\bibinfo {year} {2006}{\natexlab{a}})},\ \bibinfo {note}
	{publisher: American Association for the Advancement of Science}\BibitemShut
	{NoStop}%
	\bibitem [{\citenamefont {Nielsen}\ \emph
		{et~al.}(2006{\natexlab{b}})\citenamefont {Nielsen}, \citenamefont {Dowling},
		\citenamefont {Gu},\ and\ \citenamefont {Doherty}}]{nielsen2006optimal}%
	\BibitemOpen
	\bibfield  {author} {\bibinfo {author} {\bibfnamefont {M.~A.}\ \bibnamefont
			{Nielsen}}, \bibinfo {author} {\bibfnamefont {M.~R.}\ \bibnamefont
			{Dowling}}, \bibinfo {author} {\bibfnamefont {M.}~\bibnamefont {Gu}},\ and\
		\bibinfo {author} {\bibfnamefont {A.~C.}\ \bibnamefont {Doherty}},\
	}\bibfield  {title} {\bibinfo {title} {Optimal control, geometry, and quantum
			computing},\ }\href@noop {} {\bibfield  {journal} {\bibinfo  {journal}
			{Physical Review A—Atomic, Molecular, and Optical Physics}\ }\textbf
		{\bibinfo {volume} {73}},\ \bibinfo {pages} {062323} (\bibinfo {year}
		{2006}{\natexlab{b}})}\BibitemShut {NoStop}%
	\bibitem [{\citenamefont {Bhattacharyya}\ \emph {et~al.}(2020)\citenamefont
		{Bhattacharyya}, \citenamefont {Nandy},\ and\ \citenamefont
		{Sinha}}]{bhattacharyya2020renormalized}%
	\BibitemOpen
	\bibfield  {author} {\bibinfo {author} {\bibfnamefont {A.}~\bibnamefont
			{Bhattacharyya}}, \bibinfo {author} {\bibfnamefont {P.}~\bibnamefont
			{Nandy}},\ and\ \bibinfo {author} {\bibfnamefont {A.}~\bibnamefont {Sinha}},\
	}\bibfield  {title} {\bibinfo {title} {Renormalized circuit complexity},\
	}\href@noop {} {\bibfield  {journal} {\bibinfo  {journal} {Physical Review
				Letters}\ }\textbf {\bibinfo {volume} {124}},\ \bibinfo {pages} {101602}
		(\bibinfo {year} {2020})}\BibitemShut {NoStop}%
	\bibitem [{\citenamefont {Perrier}\ \emph {et~al.}(2020)\citenamefont
		{Perrier}, \citenamefont {Tao},\ and\ \citenamefont
		{Ferrie}}]{perrier2020quantum}%
	\BibitemOpen
	\bibfield  {author} {\bibinfo {author} {\bibfnamefont {E.}~\bibnamefont
			{Perrier}}, \bibinfo {author} {\bibfnamefont {D.}~\bibnamefont {Tao}},\ and\
		\bibinfo {author} {\bibfnamefont {C.}~\bibnamefont {Ferrie}},\ }\bibfield
	{title} {\bibinfo {title} {Quantum geometric machine learning for quantum
			circuits and control},\ }\href@noop {} {\bibfield  {journal} {\bibinfo
			{journal} {New Journal of Physics}\ }\textbf {\bibinfo {volume} {22}},\
		\bibinfo {pages} {103056} (\bibinfo {year} {2020})}\BibitemShut {NoStop}%
	\bibitem [{\citenamefont {Carlini}\ \emph {et~al.}(2007)\citenamefont
		{Carlini}, \citenamefont {Hosoya}, \citenamefont {Koike},\ and\ \citenamefont
		{Okudaira}}]{carlini2007time}%
	\BibitemOpen
	\bibfield  {author} {\bibinfo {author} {\bibfnamefont {A.}~\bibnamefont
			{Carlini}}, \bibinfo {author} {\bibfnamefont {A.}~\bibnamefont {Hosoya}},
		\bibinfo {author} {\bibfnamefont {T.}~\bibnamefont {Koike}},\ and\ \bibinfo
		{author} {\bibfnamefont {Y.}~\bibnamefont {Okudaira}},\ }\bibfield  {title}
	{\bibinfo {title} {Time-optimal unitary operations},\ }\href@noop {}
	{\bibfield  {journal} {\bibinfo  {journal} {Physical Review A—Atomic,
				Molecular, and Optical Physics}\ }\textbf {\bibinfo {volume} {75}},\ \bibinfo
		{pages} {042308} (\bibinfo {year} {2007})}\BibitemShut {NoStop}%
	\bibitem [{\citenamefont {Wang}\ \emph {et~al.}(2015)\citenamefont {Wang},
		\citenamefont {Allegra}, \citenamefont {Jacobs}, \citenamefont {Lloyd},
		\citenamefont {Lupo},\ and\ \citenamefont {Mohseni}}]{wang_quantum_2015}%
	\BibitemOpen
	\bibfield  {author} {\bibinfo {author} {\bibfnamefont {X.}~\bibnamefont
			{Wang}}, \bibinfo {author} {\bibfnamefont {M.}~\bibnamefont {Allegra}},
		\bibinfo {author} {\bibfnamefont {K.}~\bibnamefont {Jacobs}}, \bibinfo
		{author} {\bibfnamefont {S.}~\bibnamefont {Lloyd}}, \bibinfo {author}
		{\bibfnamefont {C.}~\bibnamefont {Lupo}},\ and\ \bibinfo {author}
		{\bibfnamefont {M.}~\bibnamefont {Mohseni}},\ }\bibfield  {title} {\bibinfo
		{title} {Quantum {Brachistochrone} {Curves} as {Geodesics}: {Obtaining}
			{Accurate} {Minimum}-{Time} {Protocols} for the {Control} of {Quantum}
			{Systems}},\ }\href {https://doi.org/10.1103/PhysRevLett.114.170501}
	{\bibfield  {journal} {\bibinfo  {journal} {Physical Review Letters}\
		}\textbf {\bibinfo {volume} {114}},\ \bibinfo {pages} {170501} (\bibinfo
		{year} {2015})},\ \bibinfo {note} {publisher: American Physical
		Society}\BibitemShut {NoStop}%
	\bibitem [{\citenamefont {Swaddle}\ \emph {et~al.}(2017)\citenamefont
		{Swaddle}, \citenamefont {Noakes}, \citenamefont {Smallbone}, \citenamefont
		{Salter},\ and\ \citenamefont {Wang}}]{swaddle2017generating}%
	\BibitemOpen
	\bibfield  {author} {\bibinfo {author} {\bibfnamefont {M.}~\bibnamefont
			{Swaddle}}, \bibinfo {author} {\bibfnamefont {L.}~\bibnamefont {Noakes}},
		\bibinfo {author} {\bibfnamefont {H.}~\bibnamefont {Smallbone}}, \bibinfo
		{author} {\bibfnamefont {L.}~\bibnamefont {Salter}},\ and\ \bibinfo {author}
		{\bibfnamefont {J.}~\bibnamefont {Wang}},\ }\bibfield  {title} {\bibinfo
		{title} {Generating three-qubit quantum circuits with neural networks},\
	}\href@noop {} {\bibfield  {journal} {\bibinfo  {journal} {Physics Letters
				A}\ }\textbf {\bibinfo {volume} {381}},\ \bibinfo {pages} {3391} (\bibinfo
		{year} {2017})}\BibitemShut {NoStop}%
	\bibitem [{\citenamefont {Swaddle}(2017)}]{swaddle2017thesis}%
	\BibitemOpen
	\bibfield  {author} {\bibinfo {author} {\bibfnamefont {M.}~\bibnamefont
			{Swaddle}},\ }{\selectlanguage {English}\emph {\bibinfo {title}
			{SubRiemannian geodesics and cubics for efficient quantum circuits}}},\ \href
	{https://doi.org/10.4225/23/59bf1e09f6cd7} {Master's thesis},\ \bibinfo
	{school} {The University of Western Australia} (\bibinfo {year}
	{2017})\BibitemShut {NoStop}%
	\bibitem [{\citenamefont {Lewis}\ \emph {et~al.}(2024)\citenamefont {Lewis},
		\citenamefont {Wiersema}, \citenamefont {Carrasquilla},\ and\ \citenamefont
		{Bose}}]{lewis_geodesic_2024}%
	\BibitemOpen
	\bibfield  {author} {\bibinfo {author} {\bibfnamefont {D.}~\bibnamefont
			{Lewis}}, \bibinfo {author} {\bibfnamefont {R.}~\bibnamefont {Wiersema}},
		\bibinfo {author} {\bibfnamefont {J.}~\bibnamefont {Carrasquilla}},\ and\
		\bibinfo {author} {\bibfnamefont {S.}~\bibnamefont {Bose}},\ }\href
	{https://doi.org/10.48550/arXiv.2401.05973} {\bibinfo {title} {Geodesic
			{Algorithm} for {Unitary} {Gate} {Design} with {Time}-{Independent}
			{Hamiltonians}}} (\bibinfo {year} {2024}),\ \bibinfo {note} {arXiv:2401.05973
		[quant-ph]}\BibitemShut {NoStop}%
	\bibitem [{\citenamefont {d’Alessandro}(2021)}]{d2021introduction}%
	\BibitemOpen
	\bibfield  {author} {\bibinfo {author} {\bibfnamefont {D.}~\bibnamefont
			{d’Alessandro}},\ }\href@noop {} {\emph {\bibinfo {title} {Introduction to
				quantum control and dynamics}}}\ (\bibinfo  {publisher} {Chapman and
		hall/CRC},\ \bibinfo {year} {2021})\BibitemShut {NoStop}%
	\bibitem [{\citenamefont {Rabitz}\ \emph {et~al.}(2004)\citenamefont {Rabitz},
		\citenamefont {Hsieh},\ and\ \citenamefont {Rosenthal}}]{rabitz2004quantum}%
	\BibitemOpen
	\bibfield  {author} {\bibinfo {author} {\bibfnamefont {H.~A.}\ \bibnamefont
			{Rabitz}}, \bibinfo {author} {\bibfnamefont {M.~M.}\ \bibnamefont {Hsieh}},\
		and\ \bibinfo {author} {\bibfnamefont {C.~M.}\ \bibnamefont {Rosenthal}},\
	}\bibfield  {title} {\bibinfo {title} {Quantum optimally controlled
			transition landscapes},\ }\href@noop {} {\bibfield  {journal} {\bibinfo
			{journal} {Science}\ }\textbf {\bibinfo {volume} {303}},\ \bibinfo {pages}
		{1998} (\bibinfo {year} {2004})}\BibitemShut {NoStop}%
	\bibitem [{\citenamefont {Rabitz}\ \emph {et~al.}(2012)\citenamefont {Rabitz},
		\citenamefont {Ho}, \citenamefont {Long}, \citenamefont {Wu},\ and\
		\citenamefont {Brif}}]{rabitz2012comment}%
	\BibitemOpen
	\bibfield  {author} {\bibinfo {author} {\bibfnamefont {H.}~\bibnamefont
			{Rabitz}}, \bibinfo {author} {\bibfnamefont {T.-S.}\ \bibnamefont {Ho}},
		\bibinfo {author} {\bibfnamefont {R.}~\bibnamefont {Long}}, \bibinfo {author}
		{\bibfnamefont {R.}~\bibnamefont {Wu}},\ and\ \bibinfo {author}
		{\bibfnamefont {C.}~\bibnamefont {Brif}},\ }\bibfield  {title} {\bibinfo
		{title} {Comment on “are there traps in quantum control landscapes?”},\
	}\href@noop {} {\bibfield  {journal} {\bibinfo  {journal} {Physical review
				letters}\ }\textbf {\bibinfo {volume} {108}},\ \bibinfo {pages} {198901}
		(\bibinfo {year} {2012})}\BibitemShut {NoStop}%
	\bibitem [{\citenamefont {Moore~Tibbetts}\ \emph {et~al.}(2012)\citenamefont
		{Moore~Tibbetts}, \citenamefont {Brif}, \citenamefont {Grace}, \citenamefont
		{Donovan}, \citenamefont {Hocker}, \citenamefont {Ho}, \citenamefont {Wu},\
		and\ \citenamefont {Rabitz}}]{tibbetts2012}%
	\BibitemOpen
	\bibfield  {author} {\bibinfo {author} {\bibfnamefont {K.~W.}\ \bibnamefont
			{Moore~Tibbetts}}, \bibinfo {author} {\bibfnamefont {C.}~\bibnamefont
			{Brif}}, \bibinfo {author} {\bibfnamefont {M.~D.}\ \bibnamefont {Grace}},
		\bibinfo {author} {\bibfnamefont {A.}~\bibnamefont {Donovan}}, \bibinfo
		{author} {\bibfnamefont {D.~L.}\ \bibnamefont {Hocker}}, \bibinfo {author}
		{\bibfnamefont {T.-S.}\ \bibnamefont {Ho}}, \bibinfo {author} {\bibfnamefont
			{R.-B.}\ \bibnamefont {Wu}},\ and\ \bibinfo {author} {\bibfnamefont
			{H.}~\bibnamefont {Rabitz}},\ }\bibfield  {title} {\bibinfo {title}
		{Exploring the tradeoff between fidelity and time optimal control of quantum
			unitary transformations},\ }\href
	{https://doi.org/10.1103/PhysRevA.86.062309} {\bibfield  {journal} {\bibinfo
			{journal} {Phys. Rev. A}\ }\textbf {\bibinfo {volume} {86}},\ \bibinfo
		{pages} {062309} (\bibinfo {year} {2012})}\BibitemShut {NoStop}%
	\bibitem [{\citenamefont {Ge}\ \emph {et~al.}(2022)\citenamefont {Ge},
		\citenamefont {Wu},\ and\ \citenamefont {Rabitz}}]{ge2022optimization}%
	\BibitemOpen
	\bibfield  {author} {\bibinfo {author} {\bibfnamefont {X.}~\bibnamefont
			{Ge}}, \bibinfo {author} {\bibfnamefont {R.-B.}\ \bibnamefont {Wu}},\ and\
		\bibinfo {author} {\bibfnamefont {H.}~\bibnamefont {Rabitz}},\ }\bibfield
	{title} {\bibinfo {title} {The optimization landscape of hybrid
			quantum--classical algorithms: From quantum control to nisq applications},\
	}\href@noop {} {\bibfield  {journal} {\bibinfo  {journal} {Annual Reviews in
				Control}\ }\textbf {\bibinfo {volume} {54}},\ \bibinfo {pages} {314}
		(\bibinfo {year} {2022})}\BibitemShut {NoStop}%
	\bibitem [{\citenamefont {Beato}\ \emph {et~al.}(2024)\citenamefont {Beato},
		\citenamefont {Patil},\ and\ \citenamefont {Bukov}}]{beato2024towards}%
	\BibitemOpen
	\bibfield  {author} {\bibinfo {author} {\bibfnamefont {N.}~\bibnamefont
			{Beato}}, \bibinfo {author} {\bibfnamefont {P.}~\bibnamefont {Patil}},\ and\
		\bibinfo {author} {\bibfnamefont {M.}~\bibnamefont {Bukov}},\ }\bibfield
	{title} {\bibinfo {title} {Towards a theory of phase transitions in quantum
			control landscapes},\ }\href@noop {} {\bibfield  {journal} {\bibinfo
			{journal} {arXiv preprint arXiv:2408.11110}\ } (\bibinfo {year}
		{2024})}\BibitemShut {NoStop}%
	\bibitem [{\citenamefont {Lewis}\ \emph {et~al.}(2025)\citenamefont {Lewis},
		\citenamefont {Wiersema}, \citenamefont {Carrasquilla},\ and\ \citenamefont
		{Bose}}]{lewis_geodesic_2025}%
	\BibitemOpen
	\bibfield  {author} {\bibinfo {author} {\bibfnamefont {D.}~\bibnamefont
			{Lewis}}, \bibinfo {author} {\bibfnamefont {R.}~\bibnamefont {Wiersema}},
		\bibinfo {author} {\bibfnamefont {J.}~\bibnamefont {Carrasquilla}},\ and\
		\bibinfo {author} {\bibfnamefont {S.}~\bibnamefont {Bose}},\ }\bibfield
	{title} {\bibinfo {title} {Geodesic algorithm for unitary gate design with
			time-independent {Hamiltonians}},\ }\href
	{https://doi.org/10.1103/PhysRevA.111.052618} {\bibfield  {journal} {\bibinfo
			{journal} {Physical Review A}\ }\textbf {\bibinfo {volume} {111}},\ \bibinfo
		{pages} {052618} (\bibinfo {year} {2025})},\ \bibinfo {note} {publisher:
		American Physical Society}\BibitemShut {NoStop}%
	\bibitem [{\citenamefont {Morgado}\ and\ \citenamefont
		{Whitlock}(2021)}]{morgado_quantum_2021}%
	\BibitemOpen
	\bibfield  {author} {\bibinfo {author} {\bibfnamefont {M.}~\bibnamefont
			{Morgado}}\ and\ \bibinfo {author} {\bibfnamefont {S.}~\bibnamefont
			{Whitlock}},\ }\bibfield  {title} {\bibinfo {title} {Quantum simulation and
			computing with {Rydberg}-interacting qubits},\ }\href
	{https://doi.org/10.1116/5.0036562} {\bibfield  {journal} {\bibinfo
			{journal} {AVS Quantum Science}\ }\textbf {\bibinfo {volume} {3}},\ \bibinfo
		{pages} {023501} (\bibinfo {year} {2021})},\ \bibinfo {note}
	{arXiv:2011.03031 [cond-mat, physics:physics, physics:quant-ph]}\BibitemShut
	{NoStop}%
	\bibitem [{\citenamefont {Adams}\ \emph {et~al.}(2019)\citenamefont {Adams},
		\citenamefont {Pritchard},\ and\ \citenamefont
		{Shaffer}}]{adams_rydberg_2019}%
	\BibitemOpen
	\bibfield  {author} {\bibinfo {author} {\bibfnamefont {C.~S.}\ \bibnamefont
			{Adams}}, \bibinfo {author} {\bibfnamefont {J.~D.}\ \bibnamefont
			{Pritchard}},\ and\ \bibinfo {author} {\bibfnamefont {J.~P.}\ \bibnamefont
			{Shaffer}},\ }\bibfield  {title} {{\selectlanguage {en}\bibinfo {title}
			{Rydberg atom quantum technologies}},\ }\href
	{https://doi.org/10.1088/1361-6455/ab52ef} {\bibfield  {journal} {\bibinfo
			{journal} {Journal of Physics B: Atomic, Molecular and Optical Physics}\
		}\textbf {\bibinfo {volume} {53}},\ \bibinfo {pages} {012002} (\bibinfo
		{year} {2019})},\ \bibinfo {note} {publisher: IOP Publishing}\BibitemShut
	{NoStop}%
	\bibitem [{\citenamefont {Kingma}\ and\ \citenamefont
		{Ba}(2017)}]{kingma_adam_2017}%
	\BibitemOpen
	\bibfield  {author} {\bibinfo {author} {\bibfnamefont {D.~P.}\ \bibnamefont
			{Kingma}}\ and\ \bibinfo {author} {\bibfnamefont {J.}~\bibnamefont {Ba}},\
	}\href {https://doi.org/10.48550/arXiv.1412.6980} {\bibinfo {title} {Adam:
			{A} {Method} for {Stochastic} {Optimization}}} (\bibinfo {year} {2017}),\
	\bibinfo {note} {arXiv:1412.6980 [cs]}\BibitemShut {NoStop}%
	\bibitem [{\citenamefont {Goodwin}\ and\ \citenamefont
		{Kuprov}(2015)}]{goodwin_auxiliary_2015}%
	\BibitemOpen
	\bibfield  {author} {\bibinfo {author} {\bibfnamefont {D.~L.}\ \bibnamefont
			{Goodwin}}\ and\ \bibinfo {author} {\bibfnamefont {I.}~\bibnamefont
			{Kuprov}},\ }\bibfield  {title} {\bibinfo {title} {Auxiliary matrix formalism
			for interaction representation transformations, optimal control, and spin
			relaxation theories},\ }\href {https://doi.org/10.1063/1.4928978} {\bibfield
		{journal} {\bibinfo  {journal} {The Journal of Chemical Physics}\ }\textbf
		{\bibinfo {volume} {143}},\ \bibinfo {pages} {084113} (\bibinfo {year}
		{2015})}\BibitemShut {NoStop}%
	\bibitem [{\citenamefont {Goodwin}\ and\ \citenamefont
		{Kuprov}(2016)}]{goodwin_modified_2016}%
	\BibitemOpen
	\bibfield  {author} {\bibinfo {author} {\bibfnamefont {D.~L.}\ \bibnamefont
			{Goodwin}}\ and\ \bibinfo {author} {\bibfnamefont {I.}~\bibnamefont
			{Kuprov}},\ }\bibfield  {title} {\bibinfo {title} {Modified
			{Newton}-{Raphson} {GRAPE} methods for optimal control of spin systems},\
	}\href {https://doi.org/10.1063/1.4949534} {\bibfield  {journal} {\bibinfo
			{journal} {The Journal of Chemical Physics}\ }\textbf {\bibinfo {volume}
			{144}},\ \bibinfo {pages} {204107} (\bibinfo {year} {2016})}\BibitemShut
	{NoStop}%
	\bibitem [{\citenamefont {Lewis}\ and\ \citenamefont
		{Wiersema}(2025)}]{our_data}%
	\BibitemOpen
	\bibfield  {author} {\bibinfo {author} {\bibfnamefont {D.}~\bibnamefont
			{Lewis}}\ and\ \bibinfo {author} {\bibfnamefont {R.}~\bibnamefont
			{Wiersema}},\ }\href@noop {} {\bibinfo {title} {{Quantum Optimal Control with
				Geodesic Pulse Engineering}}} (\bibinfo {year} {2025}),\ \bibinfo {note}
	{\url{https://github.com/dyylan/geodesic_control}}\BibitemShut {NoStop}%
	\bibitem [{\citenamefont {Nogueira}(14  )}]{bayesian_optimization_python}%
	\BibitemOpen
	\bibfield  {author} {\bibinfo {author} {\bibfnamefont {F.}~\bibnamefont
			{Nogueira}},\ }\href
	{https://github.com/bayesian-optimization/BayesianOptimization} {\bibinfo
		{title} {{Bayesian Optimization}: Open source constrained global optimization
			tool for {Python}}} (\bibinfo {year} {2014--})\BibitemShut {NoStop}%
	\bibitem [{\citenamefont {Bond}\ \emph {et~al.}(2024)\citenamefont {Bond},
		\citenamefont {Safavi-Naini},\ and\ \citenamefont
		{Minář}}]{bond_fast_2024}%
	\BibitemOpen
	\bibfield  {author} {\bibinfo {author} {\bibfnamefont {L.~J.}\ \bibnamefont
			{Bond}}, \bibinfo {author} {\bibfnamefont {A.}~\bibnamefont {Safavi-Naini}},\
		and\ \bibinfo {author} {\bibfnamefont {J.}~\bibnamefont {Minář}},\
	}\bibfield  {title} {\bibinfo {title} {Fast {Quantum} {State} {Preparation}
			and {Bath} {Dynamics} {Using} {Non}-{Gaussian} {Variational} {Ansatz} and
			{Quantum} {Optimal} {Control}},\ }\href
	{https://doi.org/10.1103/PhysRevLett.132.170401} {\bibfield  {journal}
		{\bibinfo  {journal} {Physical Review Letters}\ }\textbf {\bibinfo {volume}
			{132}},\ \bibinfo {pages} {170401} (\bibinfo {year} {2024})},\ \bibinfo
	{note} {publisher: American Physical Society}\BibitemShut {NoStop}%
	\bibitem [{\citenamefont {Wiersema}\ \emph {et~al.}(2025)\citenamefont
		{Wiersema}, \citenamefont {Kemper}, \citenamefont {Bakalov},\ and\
		\citenamefont {Killoran}}]{wiersema2025horizontal}%
	\BibitemOpen
	\bibfield  {author} {\bibinfo {author} {\bibfnamefont {R.}~\bibnamefont
			{Wiersema}}, \bibinfo {author} {\bibfnamefont {A.~F.}\ \bibnamefont
			{Kemper}}, \bibinfo {author} {\bibfnamefont {B.~N.}\ \bibnamefont
			{Bakalov}},\ and\ \bibinfo {author} {\bibfnamefont {N.}~\bibnamefont
			{Killoran}},\ }\bibfield  {title} {\bibinfo {title} {Geometric quantum
			machine learning with horizontal quantum gates},\ }\href
	{https://doi.org/10.1103/PhysRevResearch.7.013148} {\bibfield  {journal}
		{\bibinfo  {journal} {Phys. Rev. Res.}\ }\textbf {\bibinfo {volume} {7}},\
		\bibinfo {pages} {013148} (\bibinfo {year} {2025})}\BibitemShut {NoStop}%
	\bibitem [{\citenamefont {Nyisomeh}\ \emph {et~al.}(2020)\citenamefont
		{Nyisomeh}, \citenamefont {Diffo}, \citenamefont {Ateuafack},\ and\
		\citenamefont {Fai}}]{nyisomeh_landauzener_2020}%
	\BibitemOpen
	\bibfield  {author} {\bibinfo {author} {\bibfnamefont {I.~F.}\ \bibnamefont
			{Nyisomeh}}, \bibinfo {author} {\bibfnamefont {J.~T.}\ \bibnamefont {Diffo}},
		\bibinfo {author} {\bibfnamefont {M.~E.}\ \bibnamefont {Ateuafack}},\ and\
		\bibinfo {author} {\bibfnamefont {L.~C.}\ \bibnamefont {Fai}},\ }\bibfield
	{title} {\bibinfo {title} {Landau–{Zener} transitions in coupled qubits:
			{Effects} of coloured noise},\ }\href
	{https://doi.org/10.1016/j.physe.2019.113744} {\bibfield  {journal} {\bibinfo
			{journal} {Physica E: Low-dimensional Systems and Nanostructures}\ }\textbf
		{\bibinfo {volume} {116}},\ \bibinfo {pages} {113744} (\bibinfo {year}
		{2020})}\BibitemShut {NoStop}%
	\bibitem [{\citenamefont {McEwen}\ \emph {et~al.}(2021)\citenamefont {McEwen},
		\citenamefont {Kafri}, \citenamefont {Chen}, \citenamefont {Atalaya},
		\citenamefont {Satzinger}, \citenamefont {Quintana}, \citenamefont {Klimov},
		\citenamefont {Sank}, \citenamefont {Gidney}, \citenamefont {Fowler},
		\citenamefont {Arute}, \citenamefont {Arya}, \citenamefont {Buckley},
		\citenamefont {Burkett}, \citenamefont {Bushnell}, \citenamefont {Chiaro},
		\citenamefont {Collins}, \citenamefont {Demura}, \citenamefont {Dunsworth},
		\citenamefont {Erickson}, \citenamefont {Foxen}, \citenamefont {Giustina},
		\citenamefont {Huang}, \citenamefont {Hong}, \citenamefont {Jeffrey},
		\citenamefont {Kim}, \citenamefont {Kechedzhi}, \citenamefont {Kostritsa},
		\citenamefont {Laptev}, \citenamefont {Megrant}, \citenamefont {Mi},
		\citenamefont {Mutus}, \citenamefont {Naaman}, \citenamefont {Neeley},
		\citenamefont {Neill}, \citenamefont {Niu}, \citenamefont {Paler},
		\citenamefont {Redd}, \citenamefont {Roushan}, \citenamefont {White},
		\citenamefont {Yao}, \citenamefont {Yeh}, \citenamefont {Zalcman},
		\citenamefont {Chen}, \citenamefont {Smelyanskiy}, \citenamefont {Martinis},
		\citenamefont {Neven}, \citenamefont {Kelly}, \citenamefont {Korotkov},
		\citenamefont {Petukhov},\ and\ \citenamefont
		{Barends}}]{mcewen_removing_2021}%
	\BibitemOpen
	\bibfield  {author} {\bibinfo {author} {\bibfnamefont {M.}~\bibnamefont
			{McEwen}}, \bibinfo {author} {\bibfnamefont {D.}~\bibnamefont {Kafri}},
		\bibinfo {author} {\bibfnamefont {Z.}~\bibnamefont {Chen}}, \bibinfo {author}
		{\bibfnamefont {J.}~\bibnamefont {Atalaya}}, \bibinfo {author} {\bibfnamefont
			{K.~J.}\ \bibnamefont {Satzinger}}, \bibinfo {author} {\bibfnamefont
			{C.}~\bibnamefont {Quintana}}, \bibinfo {author} {\bibfnamefont {P.~V.}\
			\bibnamefont {Klimov}}, \bibinfo {author} {\bibfnamefont {D.}~\bibnamefont
			{Sank}}, \bibinfo {author} {\bibfnamefont {C.}~\bibnamefont {Gidney}},
		\bibinfo {author} {\bibfnamefont {A.~G.}\ \bibnamefont {Fowler}}, \bibinfo
		{author} {\bibfnamefont {F.}~\bibnamefont {Arute}}, \bibinfo {author}
		{\bibfnamefont {K.}~\bibnamefont {Arya}}, \bibinfo {author} {\bibfnamefont
			{B.}~\bibnamefont {Buckley}}, \bibinfo {author} {\bibfnamefont
			{B.}~\bibnamefont {Burkett}}, \bibinfo {author} {\bibfnamefont
			{N.}~\bibnamefont {Bushnell}}, \bibinfo {author} {\bibfnamefont
			{B.}~\bibnamefont {Chiaro}}, \bibinfo {author} {\bibfnamefont
			{R.}~\bibnamefont {Collins}}, \bibinfo {author} {\bibfnamefont
			{S.}~\bibnamefont {Demura}}, \bibinfo {author} {\bibfnamefont
			{A.}~\bibnamefont {Dunsworth}}, \bibinfo {author} {\bibfnamefont
			{C.}~\bibnamefont {Erickson}}, \bibinfo {author} {\bibfnamefont
			{B.}~\bibnamefont {Foxen}}, \bibinfo {author} {\bibfnamefont
			{M.}~\bibnamefont {Giustina}}, \bibinfo {author} {\bibfnamefont
			{T.}~\bibnamefont {Huang}}, \bibinfo {author} {\bibfnamefont
			{S.}~\bibnamefont {Hong}}, \bibinfo {author} {\bibfnamefont {E.}~\bibnamefont
			{Jeffrey}}, \bibinfo {author} {\bibfnamefont {S.}~\bibnamefont {Kim}},
		\bibinfo {author} {\bibfnamefont {K.}~\bibnamefont {Kechedzhi}}, \bibinfo
		{author} {\bibfnamefont {F.}~\bibnamefont {Kostritsa}}, \bibinfo {author}
		{\bibfnamefont {P.}~\bibnamefont {Laptev}}, \bibinfo {author} {\bibfnamefont
			{A.}~\bibnamefont {Megrant}}, \bibinfo {author} {\bibfnamefont
			{X.}~\bibnamefont {Mi}}, \bibinfo {author} {\bibfnamefont {J.}~\bibnamefont
			{Mutus}}, \bibinfo {author} {\bibfnamefont {O.}~\bibnamefont {Naaman}},
		\bibinfo {author} {\bibfnamefont {M.}~\bibnamefont {Neeley}}, \bibinfo
		{author} {\bibfnamefont {C.}~\bibnamefont {Neill}}, \bibinfo {author}
		{\bibfnamefont {M.}~\bibnamefont {Niu}}, \bibinfo {author} {\bibfnamefont
			{A.}~\bibnamefont {Paler}}, \bibinfo {author} {\bibfnamefont
			{N.}~\bibnamefont {Redd}}, \bibinfo {author} {\bibfnamefont {P.}~\bibnamefont
			{Roushan}}, \bibinfo {author} {\bibfnamefont {T.~C.}\ \bibnamefont {White}},
		\bibinfo {author} {\bibfnamefont {J.}~\bibnamefont {Yao}}, \bibinfo {author}
		{\bibfnamefont {P.}~\bibnamefont {Yeh}}, \bibinfo {author} {\bibfnamefont
			{A.}~\bibnamefont {Zalcman}}, \bibinfo {author} {\bibfnamefont
			{Y.}~\bibnamefont {Chen}}, \bibinfo {author} {\bibfnamefont {V.~N.}\
			\bibnamefont {Smelyanskiy}}, \bibinfo {author} {\bibfnamefont {J.~M.}\
			\bibnamefont {Martinis}}, \bibinfo {author} {\bibfnamefont {H.}~\bibnamefont
			{Neven}}, \bibinfo {author} {\bibfnamefont {J.}~\bibnamefont {Kelly}},
		\bibinfo {author} {\bibfnamefont {A.~N.}\ \bibnamefont {Korotkov}}, \bibinfo
		{author} {\bibfnamefont {A.~G.}\ \bibnamefont {Petukhov}},\ and\ \bibinfo
		{author} {\bibfnamefont {R.}~\bibnamefont {Barends}},\ }\bibfield  {title}
	{{\selectlanguage {en}\bibinfo {title} {Removing leakage-induced correlated
				errors in superconducting quantum error correction}},\ }\href
	{https://doi.org/10.1038/s41467-021-21982-y} {\bibfield  {journal} {\bibinfo
			{journal} {Nature Communications}\ }\textbf {\bibinfo {volume} {12}},\
		\bibinfo {pages} {1761} (\bibinfo {year} {2021})},\ \bibinfo {note}
	{publisher: Nature Publishing Group}\BibitemShut {NoStop}%
	\bibitem [{\citenamefont {Ellert-Beck}\ and\ \citenamefont
		{Ge}(2024)}]{ellert-beck_power-optimized_2024}%
	\BibitemOpen
	\bibfield  {author} {\bibinfo {author} {\bibfnamefont {L.}~\bibnamefont
			{Ellert-Beck}}\ and\ \bibinfo {author} {\bibfnamefont {W.}~\bibnamefont
			{Ge}},\ }\href {https://doi.org/10.48550/arXiv.2412.17789} {\bibinfo {title}
		{Power-optimized amplitude modulation for robust trapped-ion entangling
			gates: a study of gate-timing errors}} (\bibinfo {year} {2024}),\ \bibinfo
	{note} {arXiv:2412.17789 [quant-ph]}\BibitemShut {NoStop}%
	\bibitem [{\citenamefont {Helgason}(1979)}]{helgason_differential_1979}%
	\BibitemOpen
	\bibfield  {author} {\bibinfo {author} {\bibfnamefont {S.}~\bibnamefont
			{Helgason}},\ }\href@noop {} {{\selectlanguage {en}\emph {\bibinfo {title}
				{Differential {Geometry}, {Lie} {Groups}, and {Symmetric} {Spaces}}}}}\
	(\bibinfo  {publisher} {Academic Press},\ \bibinfo {year} {1979})\ \bibinfo
	{note} {google-Books-ID: DWGvsa6bcuMC}\BibitemShut {NoStop}%
	\bibitem [{\citenamefont {Bradbury}\ \emph {et~al.}(2018)\citenamefont
		{Bradbury}, \citenamefont {Frostig}, \citenamefont {Hawkins}, \citenamefont
		{Johnson}, \citenamefont {Leary}, \citenamefont {Maclaurin}, \citenamefont
		{Necula}, \citenamefont {Paszke}, \citenamefont {Vander{P}las}, \citenamefont
		{Wanderman-{M}ilne},\ and\ \citenamefont {Zhang}}]{jax2018github}%
	\BibitemOpen
	\bibfield  {author} {\bibinfo {author} {\bibfnamefont {J.}~\bibnamefont
			{Bradbury}}, \bibinfo {author} {\bibfnamefont {R.}~\bibnamefont {Frostig}},
		\bibinfo {author} {\bibfnamefont {P.}~\bibnamefont {Hawkins}}, \bibinfo
		{author} {\bibfnamefont {M.~J.}\ \bibnamefont {Johnson}}, \bibinfo {author}
		{\bibfnamefont {C.}~\bibnamefont {Leary}}, \bibinfo {author} {\bibfnamefont
			{D.}~\bibnamefont {Maclaurin}}, \bibinfo {author} {\bibfnamefont
			{G.}~\bibnamefont {Necula}}, \bibinfo {author} {\bibfnamefont
			{A.}~\bibnamefont {Paszke}}, \bibinfo {author} {\bibfnamefont
			{J.}~\bibnamefont {Vander{P}las}}, \bibinfo {author} {\bibfnamefont
			{S.}~\bibnamefont {Wanderman-{M}ilne}},\ and\ \bibinfo {author}
		{\bibfnamefont {Q.}~\bibnamefont {Zhang}},\ }\href
	{http://github.com/google/jax} {\bibinfo {title} {{JAX}: composable
			transformations of {P}ython+{N}um{P}y programs}} (\bibinfo {year}
	{2018})\BibitemShut {NoStop}%
	\bibitem [{\citenamefont {Haber}()}]{haber_notes_nodate}%
	\BibitemOpen
	\bibfield  {author} {\bibinfo {author} {\bibfnamefont {H.~E.}\ \bibnamefont
			{Haber}},\ }\bibfield  {title} {{\selectlanguage {en}\bibinfo {title} {Notes
				on the {Matrix} {Exponential} and {Logarithm}}},\ }\href@noop {} {\
	}\BibitemShut {NoStop}%
	\bibitem [{\citenamefont {Armijo}(1966)}]{armijo1966minimization}%
	\BibitemOpen
	\bibfield  {author} {\bibinfo {author} {\bibfnamefont {L.}~\bibnamefont
			{Armijo}},\ }\bibfield  {title} {\bibinfo {title} {Minimization of functions
			having lipschitz continuous first partial derivatives},\ }\href@noop {}
	{\bibfield  {journal} {\bibinfo  {journal} {Pacific Journal of mathematics}\
		}\textbf {\bibinfo {volume} {16}},\ \bibinfo {pages} {1} (\bibinfo {year}
		{1966})}\BibitemShut {NoStop}%
	\bibitem [{\citenamefont {Banerjee}\ \emph {et~al.}(1985)\citenamefont
		{Banerjee}, \citenamefont {Adams}, \citenamefont {Simons},\ and\
		\citenamefont {Shepard}}]{banerjee1985search}%
	\BibitemOpen
	\bibfield  {author} {\bibinfo {author} {\bibfnamefont {A.}~\bibnamefont
			{Banerjee}}, \bibinfo {author} {\bibfnamefont {N.}~\bibnamefont {Adams}},
		\bibinfo {author} {\bibfnamefont {J.}~\bibnamefont {Simons}},\ and\ \bibinfo
		{author} {\bibfnamefont {R.}~\bibnamefont {Shepard}},\ }\bibfield  {title}
	{\bibinfo {title} {Search for stationary points on surfaces},\ }\href@noop {}
	{\bibfield  {journal} {\bibinfo  {journal} {The Journal of Physical
				Chemistry}\ }\textbf {\bibinfo {volume} {89}},\ \bibinfo {pages} {52}
		(\bibinfo {year} {1985})}\BibitemShut {NoStop}%
	\bibitem [{\citenamefont {Boyd}\ and\ \citenamefont
		{Vandenberghe}(2004)}]{boyd2004convex}%
	\BibitemOpen
	\bibfield  {author} {\bibinfo {author} {\bibfnamefont {S.~P.}\ \bibnamefont
			{Boyd}}\ and\ \bibinfo {author} {\bibfnamefont {L.}~\bibnamefont
			{Vandenberghe}},\ }\href {https://doi.org/10.1017/cbo9780511804441} {\emph
		{\bibinfo {title} {Convex optimization}}}\ (\bibinfo  {publisher} {Cambridge
		university press},\ \bibinfo {year} {2004})\BibitemShut {NoStop}%
	\bibitem [{\citenamefont {Frazier}(2018)}]{frazier_tutorial_2018}%
	\BibitemOpen
	\bibfield  {author} {\bibinfo {author} {\bibfnamefont {P.~I.}\ \bibnamefont
			{Frazier}},\ }\href {https://doi.org/10.48550/arXiv.1807.02811} {\bibinfo
		{title} {A {Tutorial} on {Bayesian} {Optimization}}} (\bibinfo {year}
	{2018}),\ \bibinfo {note} {arXiv:1807.02811 [stat]}\BibitemShut {NoStop}%
	\bibitem [{\citenamefont {Srinivas}\ \emph {et~al.}(2010)\citenamefont
		{Srinivas}, \citenamefont {Krause}, \citenamefont {Kakade},\ and\
		\citenamefont {Seeger}}]{srinivas_gaussian_2010}%
	\BibitemOpen
	\bibfield  {author} {\bibinfo {author} {\bibfnamefont {N.}~\bibnamefont
			{Srinivas}}, \bibinfo {author} {\bibfnamefont {A.}~\bibnamefont {Krause}},
		\bibinfo {author} {\bibfnamefont {S.}~\bibnamefont {Kakade}},\ and\ \bibinfo
		{author} {\bibfnamefont {M.}~\bibnamefont {Seeger}},\ }\bibfield  {title}
	{\bibinfo {title} {Gaussian process optimization in the bandit setting: no
			regret and experimental design},\ }in\ \href@noop {} {\emph {\bibinfo
			{booktitle} {Proceedings of the 27th {International} {Conference} on
				{International} {Conference} on {Machine} {Learning}}}},\ \bibinfo {series
		and number} {{ICML}'10}\ (\bibinfo  {publisher} {Omnipress},\ \bibinfo
	{address} {Madison, WI, USA},\ \bibinfo {year} {2010})\ pp.\ \bibinfo {pages}
	{1015--1022}\BibitemShut {NoStop}%
\end{thebibliography}
\end{document}